\documentclass[prx,floatfix,superscriptaddress,longbibliography,nofootinbib,notitlepage,twocolumn]{revtex4-1}

\usepackage{amssymb,amsmath,amstext,amsthm,mathtools}
\usepackage{url}
\usepackage{graphicx}
\usepackage{epstopdf}
\usepackage{color}
\usepackage{bm}
\usepackage{appendix}
\usepackage[T1]{fontenc}
\usepackage{bbold}
\usepackage{bbm}
\usepackage{latexsym}
\usepackage[colorlinks=true,citecolor=blue,linkcolor=magenta]{hyperref}
\usepackage{float}
\usepackage{verbatim}
\usepackage{lipsum}
\usepackage{xcolor}
\usepackage{natbib}
\usepackage{soul}
\usepackage{multirow}

\renewcommand{\l}{\left}
\renewcommand{\r}{\right}

\newcommand{\black}[1]{{\color{black}{#1}}}

\newcommand{\dd}{\text{d}}

\begin{document}

\title{Challenges for quantum computation of nonlinear dynamical systems using linear representations}

\author{Yen Ting Lin}
\affiliation{
Information Sciences Group (CCS-3), Computer, Computational and Statistical Sciences Division, Los Alamos National Laboratory, Los Alamos, NM 87545, USA.}

\author{Robert B.~Lowrie}
\affiliation{Computational Physics and Methods Group (CCS-2), Computer, Computational and Statistical Sciences Division, Los Alamos National Laboratory, Los Alamos, NM 87545, USA}

\author{Denis Aslangil}
\affiliation{Department of Aerospace Engineering \& Mechanics, University of Alabama, Tuscaloosa, AL 35487, USA}

\author{Yi\u{g}it Suba\c{s}\i}
\affiliation{
Information Sciences Group (CCS-3), Computer, Computational and Statistical Sciences Division, Los Alamos National Laboratory, Los Alamos, NM 87545, USA.}

\author{Andrew T.~Sornborger}
\affiliation{
Information Sciences Group (CCS-3), Computer, Computational and Statistical Sciences Division, Los Alamos National Laboratory, Los Alamos, NM 87545, USA.}
\affiliation{Correspondence should be sent to sornborg@lanl.gov}

\date{\today}

\begin{abstract}
\noindent
A number of recent studies have proposed that linear representations are appropriate for solving nonlinear dynamical systems with quantum computers, which fundamentally act linearly on a wave function in a Hilbert space. Linear representations, such as the Koopman representation and Koopman von Neumann mechanics, have regained attention from the dynamical-systems research community. Here, we aim to present a unified theoretical framework, currently missing in the literature, with which one can compare and relate existing methods, their conceptual basis, and their representations. We also aim to show that, despite the fact that quantum simulation of nonlinear classical systems may be possible with such linear representations, a necessary projection into a feasible finite-dimensional space will in practice eventually induce numerical artifacts which can be hard to eliminate or even control.
As a result, a practical, reliable and accurate way to use quantum computation for solving general nonlinear dynamical systems is still an open problem. 
\end{abstract}

\maketitle

\section{Introduction}

Since its inception in the early 1980's \cite{benioff1980computer,feynman2018simulating}, quantum simulation has been a major focus of algorithmic research \cite{lloyd1996universal,yoshida1990construction,boghosian1998simulating,sornborger1999higher,berry2015simulating,low2019hamiltonian}. Via their ability to encode one quantum system in another, quantum simulation methods use exponentially fewer resources, relative to a classical computer. The most recent methods \cite{berry2015simulating,low2019hamiltonian} additionally 
provide an exponential improvement in simulation precision relative to earlier Trotterization methods, making them suitable for repeated use as a simulation module in a larger circuit.

Quantum algorithms have also been put forward to solve linear systems of equations \cite{harrow2009quantum,Amb12,CKS17,subasi2019quantum,AN19} and some of these have been used to efficiently solve systems of {\it linear} classical (non-quantum) physical systems \cite{berry2014HighorderQuantumAlgorithm,berry2017quantum,costa2019quantum,cao2013quantum,sinha2010quantum,engel2019quantum}.

Initial work attempting the integration of {\it nonlinear}, classical differential equations resulted in inefficient quantum algorithms requiring exponential resources \cite{leyton2008quantum}. However, it was recently realized that, for the nonlinear case, methods for lifting nonlinear equations to an infinite-dimensional Hilbert space, such as those using Koopman--von Neumann mechanics and the Koopman representation discussed below, were the appropriate approach \cite{joseph2020KoopmanNeumannApproach,dodin2021applications,giannakis2020quantum,liu2021efficient,lloyd2020quantum,engel2021linear}.

The apparently disparate approaches that have been taken towards the lifting of nonlinear classical systems to a Hilbert space have involved using truncated Carleman linearization \cite{liu2021efficient}, simulating the dynamics of the nonlinear Schr\"{o}dinger equation \cite{lloyd2020quantum}, and the Koopman--von Neumann formulation of classical systems \cite{joseph2020KoopmanNeumannApproach}.

It is important to note that one's expectations of efficient quantum algorithms for linearly-lifted nonlinear classical systems must be tempered somewhat relative to what one would expect from quantum algorithms for quantum systems. Due to the lack of a `sign problem' for linearly lifted nonlinear systems \cite{loh1990sign,van1994fixed,troyer2005computational}, efficient classical Monte Carlo methods, such as $n$-body simulations for molecular dynamics \cite{binney2011galactic} and importance sampling \cite{neal2001annealed}, exist for nonlinear dynamics, whereas they do not, in general, for quantum systems. Although exponential improvements have been claimed for some nonlinear systems \cite{lloyd2020quantum,liu2021efficient}, much of the exponential improvement only exists relative to previous algorithms \cite{leyton2008quantum} that were exponentially worse than classical algorithms. 

The subclass of algorithms for which rigorous bounds have been found \cite{liu2021efficient} excludes almost all of the interesting models which are actively studied in the field of nonlinear dynamical systems, for example ODEs like Lorentz `63 and `96 chaotic systems, the Van der Pol oscillator, which we investigate, the Hodgkin-Huxley equations, Lotka-Volterra and Rosenzweig-MacArthur equations, and general PDEs like the non-dissipative Burgers’ equation, the Kuramoto-Sivashinsky equation, and the Navier-Stokes equation. The conditions on the Carleman method presented in \cite{liu2021efficient} and for arbitrary representations in \cite{engel2021linear} are so restrictive that the method cannot be used to solve these equations in the strongly nonlinear domains of interest. Below, we provide evidence that if one uses a construct similar to \cite{liu2021efficient}, the system exhibits instability in finite time.

One important efficiency that may be found with a quantum algorithmic approach to nonlinear classical systems is the encoding of classical information in the quantum amplitudes of a quantum register. For instance, in \cite{liu2021efficient}, the velocity as a function of grid location is encoded in the amplitudes of a quantum state of $n$ qubits, allowing $2^n$ grid locations to be encoded in a logarithmic number of qubits. Similar spatial encoding efficiencies would be expected for any nonlinear system where a dependent variable can be encoded in a quantum register \cite{engel2019quantum}. Nonetheless, an important further step that is necessary for end-to-end implementation of these algorithms is to sample the information in the quantum state that is output \cite{montanaro2016quantum,liu2021efficient}. This can be challenging and, in general, results in no superpolynomial advantage \cite{montanaro2016quantum}. One encouraging result is that, in some cases, for instance $n$-body systems, a high-order polynomial advantage may be found \cite{montanaro2016quantum}

In this article, we give an overview of methods that allow one to represent nonlinear systems via linear systems and discuss their limitations. These limitations are independent of whether a quantum or classical computer is used to solve the linear system. Thus, our conclusions are about the nature of the approximations that are made in order to transform the problem into a form suitable for solution with a quantum computer. We refer to two simple, low dimensional (1 and 2) systems of nonlinear equations in order to demonstrate our points. In particular, we show that even for such simple systems, the ability of the linear system to represent the nonlinear system is limited in various ways. We believe these features are not artifacts of low dimensionality, and thus should be kept in mind whenever QC is considered as an option for solving a nonlinear system.

The remainder of this manuscript is organized as follows. In Sec.~\ref{sec:DynamicalSystems}, 
we provide a high-level road map that unifies the various recent proposals for representing classical systems in a Hilbert space. With this framework, one can classify and juxtapose different approaches for using quantum computation to solve classical nonlinear dynamical systems.~\footnote{Here, by `solve', we mean to output a quantum state that efficiently encodes the classical solution vector. Thus approaches based on classical encodings such as in Ref.~\cite{Xu2018Turbulent} are not covered here.
We note that this state will, necessarily, need to be 
sampled many times to extract the desired information.} For instance, we show that two of the existing proposals, using Carleman linearization and using Koopman von Neumann mechanics, live in dual spaces. The former is a Koopman representation, and the latter, despite its confusing name, is a Perron--Frobenius representation. For both the Perron--Frobenius and Koopman representations, one can obtain a linear representation for any nonlinear dynamical system by trading finite for infinite dimensionality. In order to perform a computation, one must then perform some form of projection to reduce the infinite dimensional space to an operable, finite-dimensional computational space. 

In Section \ref{sec:quantumComputation}, we use numerical simulations to demonstrate potential pitfalls of such finite-dimensional projections, which invariably introduce errors. Since quantum algorithms attempt to solve the linear problem that results from this projection, even a hypothetical, perfect implementation  will have limitations. For Carleman linearization, we show that the projection often leads to uncontrolled errors, with more detail of these errors given in Appendix \ref{app:carleman}. For Koopman--von Neumann mechanics, a spatial discretization can lead to Gibbs-like phenomena which eventually destroy crucial informative statistics. We also examine the possibility of directly solving the Liouville equation, which is the root of Koopman von Neumann mechanics. We have found that discretized Liouville equations, constructed using methods for hyperbolic conservation laws, inevitably introduce intrinsic noise. While intrinsic noise can help stabilize the Gibbs-phenomena observed in Koopman--von Neumann mechanics, it can also induce other artifacts, such as stochastic phase diffusion. Our observation suggests that an optimal numerical method could strongly depend on dynamical features of the specific nonlinear problem of interest.

Finally, we provide a discussion in Sec.~\ref{sec:discussion} to summarize our viewpoint and some of the implications of our results.

\section{Dynamical systems}\label{sec:DynamicalSystems}

\subsection{Phase-space representation of dynamical systems} \label{sec:phase-space}
Throughout this manuscript, we consider a classical nonlinear, deterministic, and autonomous dynamical system
\begin{equation}
    \frac{\dd x(t)}{\dd t}  = F\l(x\l(t\r)\r), 
    \label{eq:ODE}
\end{equation}
where a state $x=\l(x_1,\ldots,x_N\r)\in \mathbb{R}^N$ is evolved by a general vector field, often referred to as the ``flow'',  $F:\mathbb{R}^N\rightarrow \mathbb{R}^N$. For simplicity, we consider the state space to be the whole of $\mathbb{R}^N$ despite that, more generally, the state space can be any compact Riemannian manifold $\mathcal{M}$ endowed with a Borel sigma algebra and a measure. We will consider locally Lipschitz continuous $F$, such that a unique $x(t)$ exists for any real $t\ge 0$. Given an initial condition, $x(0):=x_0$, the formal solution of the system at any future time $t\ge 0$ can be expressed as 
\begin{equation}
    \varphi(x_0,t) := x(t;x_0) = \int_0^t F\l(x\l(\tau\r)\r) \dd \tau + x_0.  \label{eq:formalSol}
\end{equation}
Given a time of evolution $T\ge 0$, the formal solution Eq.~\eqref{eq:formalSol} defines the nonlinear flowmap $\varphi(\cdot,T): \mathbb{R}^N\rightarrow \mathbb{R}^N$ which maps any initial condition $x(0) \in \mathbb{R}^N$ at time $0$ to its future solution at time $T$. 

\begin{figure*}[!t]
\centering
    \includegraphics[width=0.92\textwidth]{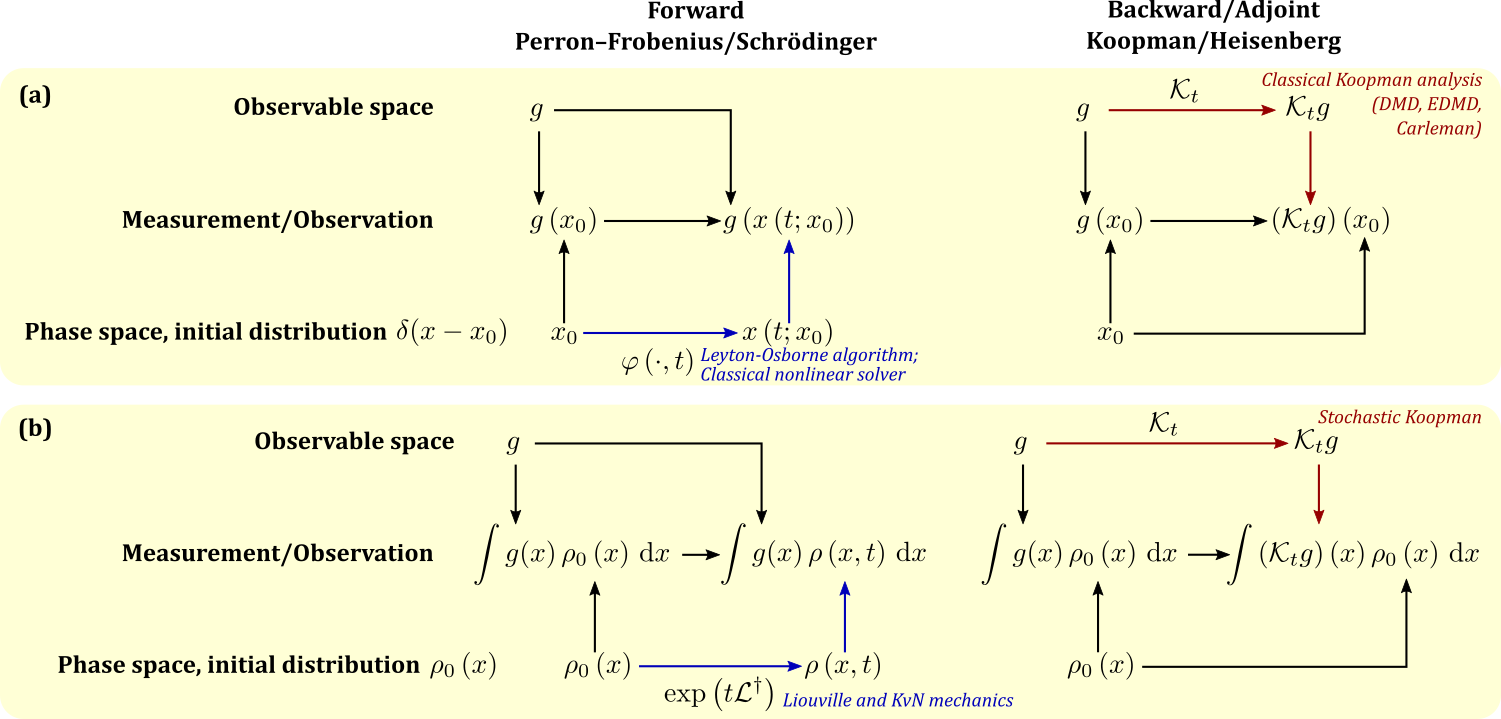}
    \caption{Schematic diagram of the relationship between the forward/Perron--Frobenius/Schr\"odinger representation and the backward/adjoint/Koopman/Heisenberg representation. In the forward picture, phase-space variables/distributions are evolving and the observables are fixed. In the backward picture, phase-space variables/distributions are fixed and the observables are evolving. (a) ``Classical'' systems given a single ($\delta$-distributed) initial condition; (b) ``Classical statistical'' systems given an initial distribution $\rho_0(x)$. }
    \label{fig:schematic}
\end{figure*}

\subsection{Perron--Frobenius representation}
Both the Koopman representation and Koopman von Neumann mechanics are closely tied to the Liouville representation of dynamical systems. In contrast to setting a single initial condition of the phase-space representation described in Sec.~\eqref{sec:phase-space}, the Liouville representation introduces a probabilistic initial condition $x(0)$ that is distributed according to $\rho_0 := \rho(x,t=0)$, and derives a partial-differential equation describing how the distribution evolves by the flow $F$:
\begin{equation}
    \frac{\partial}{\partial t} \rho(x,t) = - \sum_{j=1}^N \frac{\partial}{\partial x_j} \l[F_j \l(x\r) \rho \l(x,t\r) \r]=: \mathcal{L}^\dagger \rho(x,t). \label{eq:Liouville}
\end{equation}
Here, we define the forward operator $\mathcal{L}^\dagger$ which generates the evolution of the probability density. It is important to emphasize that the only probabilistic aspect of the system is its initial condition, given by the probability distribution, $\rho_0$. This initial distribution is propagated by the deterministic flow, $F$, to a later time, $\rho(x,t)$. Below, we denote the probability density at time $t$ as $\rho_t(x) := \rho(x,t)$. In the limit that $\rho_0$ approaches a Dirac $\delta$-distribution, the solution evolves as a $\delta$-distribution in the phase space $\mathbb{R}^N$ along the deterministic trajectory Eq.~\eqref{eq:formalSol}, and thus the Liouville representation is a generalization of the phase-space representation in Sec.~\ref{sec:phase-space}. 

Note that the Liouville representation is linear in $\rho$: suppose $\rho^{(1)}$ and $\rho^{(2)}$ both satisfy Eq.~\eqref{eq:Liouville}, their linear combination $\alpha \rho^{(1)} + \beta \rho^{(t)}$, $\alpha$, $\beta\in \mathbb{R}$ also satisfies Eq.~\eqref{eq:Liouville}. The linearity and lack of the ``sign problem'' that plagues Quantum Monte Carlo \cite{loh1990sign,van1994fixed,troyer2005computational}, leads to efficient Monte Carlo sampling for integrating \eqref{eq:Liouville}: one first samples $N_\text{MC}$ initial conditions from $\rho_0$, then independently propagates the deterministic trajectories to $t\ge 0$ by Eq.~\eqref{eq:ODE} to obtain $N_\text{MC}$ samples of $\rho_t$.

We finally remark that when the dynamics are non-deterministic, the Liouville representation can be generalized to the so-called Perron--Frobenius representation \cite{klus2018DataDrivenModelReduction}. For example, for the dynamical system Eq.~\eqref{eq:ODE}, subject to isotropic Gaussian white noise, the Perron--Frobenius representation of the stochastic dynamical system is described by the Fokker--Planck Equation:
\begin{equation}
    \frac{\partial}{\partial t} \rho(x,t) = - \nabla_x\cdot \l[F \l(x\r) \rho \l(x,t\r) \r] + \frac{D}{2} \nabla_x^2 \l[\rho (x, t)\r], \label{eq:FP}
\end{equation}
where $D$ is the isotropic diffusion coefficient. In this case, we identify the forward generator $\mathcal{L}^\dagger=-\nabla_x F(x) + D \nabla^2_x /2$, where the differential operator $\nabla_x$ acts on everything to its right.  

\subsection{Koopman representation}

In contrast to the Liouville and Perron--Frobenius representations, the other side of the coin is the Koopman representation \cite{Koopman315,Koopman255,mezic2005SpectralPropertiesDynamical}. Instead of describing the evolution of the probability density in phase space, the Koopman representation aims to describe how ``observables'' evolve under the same dynamics. An observable is defined as a real- (or complex-) valued function, $g:\mathbb{R}^N\rightarrow \mathbb{R}$ (or $\mathbb{C}$), which maps the phase-space configuration $x$ to $g(x)$. In addition, an observable must be an $L^2\l(\mathbb{R}^N, \dd \mu\r)$ function, meaning that it is square-integrable with respect to a specified measure $\dd \mu$:
\begin{equation}
    \int_{\mathbb{R}^N} \l\vert g(x) \r\vert^2 \dd \mu(x) < \infty. 
\end{equation}
In the Koopman representation, we will strictly consider the measure $\dd \mu$ to be a proper probability density with normalization, $\int_{\mathbb{R}^N} \dd \mu = 1$. 
This technical condition ensures that for all observables of interest, $L^2\l(\mathbb{R}^N, \dd \mu\r)$, also implies $L^1\l(\mathbb{R}^N, \dd \mu\r)$, established by a simple application of H\"older's inequality.

The central object in Koopman representation theory is a \emph{finite-time Koopman operator} $\mathcal{K}_t$ which maps an $L^2\l(\mathbb{R}^N, \rho_t (x) \, \dd x\r)$ function, $g$, to an $L^2\l(\mathbb{R}^N, \rho_0 (x) \, \dd x\r)$ function $\mathcal{K}_t g$, such that
\begin{equation}
   \int_{\mathbb{R}^N} \l(\mathcal{K}_t g\r) \rho_0(x)\, \dd x = \int_{\mathbb{R}^N} g(x) \rho_t(x) \, \dd x. \label{eq:finite-timeKoopman}
\end{equation}
That is, given an observable $g$, the expectation value of the observable against the \emph{evolving}  density $\rho_t$ is identical to the expectation value of $\mathcal{K}_t g$ against the \emph{initial}  density $\rho_0$.
In the above equation, $\mathcal{K}_t g $ is integrable with respect to the initial distribution $\rho_0$, but $g$ is integrable against the later distribution, $\rho_t(\cdot)$, propagated by the dynamics. 

The $L^2$-integrability ensures a well-defined inner product operation between two $L^2\l(\mathbb{R}^N, \dd \mu\r)$ functions $g_1$ and $g_2$:
\begin{equation}
    \l\langle g_1, g_2\r\rangle_\mu := \int_{\mathbb{R}^N} g_1\l(x\r) \, g_2\l(x\r) \, \dd \mu. \label{eq:innerProduct}
\end{equation}
As seen below, this inner product plays an important role in the Koopman representation. We will interchangeably use the Dirac bra-ket notation, $\l\langle g_1 \large \vert g_2 \r\rangle_\mu = \l\langle g_1, g_2 \r\rangle_\mu$. 

A schematic diagram Fig.~\eqref{fig:schematic}(a) is commonly presented in the Koopman literature for a $\delta$-distributed initial condition; i.e., $x_0$ is a point in $\mathbb{R}^N$. Figure \eqref{fig:schematic}(b) shows the slight generalization when we consider an initial distribution $\rho_0$ propagated by the nonlinear flow $F$. Importantly, \emph{the mapped observable $\mathcal{K}_t g$ depends on the initial condition $x_0$ in Fig.~\eqref{fig:schematic}(a), and is square-integrable with respect to the measure} $\rho_0(x)\, \dd x$ \emph{in Fig.~\ref{fig:schematic}(b)}.  Regardless of $t$, $\mathcal{K}_t$ always ``pulls back'' the function $g$, which is evaluated at the time $t\ge0$ and integrable with respect to an \emph{evolving} distribution $\rho_t$, to a function which is evaluated at the initial time $t=0$ with respect to a \emph{fixed distribution} $\rho_0$. 

The Koopman representation is linear even for those dynamical systems with nonlinear flow $F$ in the phase-space representation. The finite-time Koopman operator $\mathcal{K}_t$ is a linear operator, that is, given functions $g_1, g_2 \in L^2\l(\mathbb{R}^N, \rho_t \l(x\r) \dd x\r)$ and arbitrary constants $\alpha, \beta\in \mathbb{R}$ or $\mathbb{C}$, $\mathcal{K}_t$ maps $\alpha g_1 + \beta g_2$ to $\alpha \l(K_t g_1\r) + \beta \l(K_t g_2\r)$. 
Note that $\mathcal{K}_{t=0}$ is an identity mapping by definition. For any function $g\in L^2\l(\mathbb{R}^N, \rho_t\l(x\r) \dd x\r)$, sending $t\downarrow 0$ implies that $g = \mathcal{K}_0 g$ must be in $L^2\l(\mathbb{R}^N, \rho_0\l(x\r) \dd x\r)$. Together with the inner product Eq.~\eqref{eq:innerProduct} with $\dd \mu := \rho_0(x) \dd x$, all $L^2\l(\mathbb{R}^N, \rho_0\l(x\r) \dd x\r)$ functions form a Hilbert space. We remark that the inner product is uniquely defined as Eq.~\eqref{eq:innerProduct}, against the \emph{initial probability density} $\rho_0\l(x\r)$. Below, we denote the inner product by $\l\langle \cdot, \cdot \r\rangle_{\rho_0}$ for brevity. 

In this Hilbert space, suppose we have a complete set of functional bases $\l\{f_j\r\}_j$ which linearly spans $L^2\l(\mathbb{R}^N, \rho_0 \l(x\r) \dd x\r)$, we can express any function $g=\sum_j \alpha_j f_j$ where $\alpha_j$'s are the coefficients. Because $\mathcal{K}_t f_j$ is another function in $L^2\l(\mathbb{R}^N, \rho_0\l(x\r) \dd x\r)$, we can express  $\mathcal{K}_t f_j $ as a linear sum of the basis functions, $\sum_{k} K_{j,k}(t) f_k$, where $K_{j,k}(t)$ are the linear coefficients characterizing the transformation. Immediately, we observe that $\mathcal{K}_t g = \sum_{j,k} \alpha_j K_{j,k}(t) f_k$---for any $g$, the finite-time Koopman operator linearly transforms the coefficients of the bases $\l\{f_j\r\}_j$, from $\l\{\alpha_j\r\}_j$ to $\l\{\sum_k \alpha_k K_{k,j} (t)  \r\}_j$. The linearity of the Koopman representation enables the possibility of integration on a quantum computer, but comes with a trade-off: the operational space can be infinite dimensional: in most nonlinear systems, the number of indices ($i,j$) can be countably or (even worse) uncountably infinite. Even though the evolution of an observable $g$ depends linearly on other observables, this dependence can potentially involve infinitely many others; we observe similar problems in moment expansion methods \cite{ale2013general,Schnoerr2015comparison} and Carleman linearizations \cite{carleman1932,kowalski1991NonlinearDynamicalSystems}: the final closure is still an open problem, see the discussion in \cite{lin2021data}. Nevertheless, in the classical (non-quantum) context, approximation methods 
take advantage of the linearity of the Koopman representation to transform the non-convex optimization problem of learning nonlinear systems into a globally convex problem, thus enabling data-driven modeling for complex and multiscale problems \cite{schmid2010DynamicModeDecomposition,williamsDataDrivenApproximation2015,PCCA2013}. 

The linear property of $\mathcal{K}_t$ motivates us to seek Koopman eigenvalue-eigenfunction pairs satisfying $\mathcal{K}_t \phi_j = e^{\lambda_j t} \phi_j$, $\lambda_j \in \mathbb{C}$ and $\phi_j \in L^2\l(\mathbb{R}^N, \rho_0\l(x\r)\, \dd x\r)$ $\forall t\ge 0$. The Koopman eigenfunctions $\{ \phi_j \}$ are thus the ``good'' functions whose evolution is invariant: one can write $\dd  \l(\mathcal{K}_t \phi_j\r) /\dd t= \lambda_j \phi_j$. From Eq.~\eqref{eq:finite-timeKoopman}, we also know that the time dependence of the expectation value of the observable $\sim e^{\lambda_j t}$. Furthermore, suppose we use these Koopman eigenfunctions as the basis functions to expand the functional space $L^2\l(\mathbb{R}^N, \rho_0 \l(x\r)\,\dd x\r)$, the linear decomposition of any function $g = \sum_{i} \beta_j \phi_j$ keeps its form under the evolution as $\mathcal{K}_t g = \sum_j \beta_j e^{\lambda_j t} \phi_j$. The Koopman eigenfunctions also characterize those invariant structures under the dynamics; for example, the leading Koopman eigenfunctions reveal basins of attraction in multistable systems \cite{schmid2010DynamicModeDecomposition,williamsDataDrivenApproximation2015}. 

Koopman showed that for Hamiltonian systems, the Koopman operators are always unitary, and thus, the Koopman eigenfunctions are mutually orthogonal, i.e, $\l\langle \phi_j, \phi_j\r\rangle_{\rho_0} = \delta_{i,j}$. Nevertheless, because the Koopman operator of a general irreversible dynamical system is not guaranteed to be unitary, its Koopman eigenfunctions do not necessarily form an orthogonal basis set with the inner product \eqref{eq:innerProduct}. Nor are they required to be properly normalized. That is, $\l\langle \phi_j, \phi_j\r\rangle_{\rho_0} \ne \delta_{i,j}$ in general, where $\delta_{i,j}$ is the Kronecker delta function. Consequently, the projection for solving the linear coefficients ($\alpha_j$ and $\beta_j$ above) often requires a proper normalization \cite{chorin2002optimal,williamsDataDrivenApproximation2015,lin2021data}. 

We finally remark that the $j$-notation ($\lambda_j$, $\phi_j$, $\sum_j$) is a stylized book-keeping notation. Since Koopman and von Neumann's time, we have known that there are systems---for example, chaotic ones---with an uncountable number of eigenvalues \cite{Koopman255,neumann1932ZurOperatorenmethodeKlassischen,mezic2005SpectralPropertiesDynamical,mezic2020SpectrumKoopmanOperator}. The spectrum of the Koopman operator of these systems has one or more continuous parts, and the functional space is an inseparable Hilbert space. In this case, we cannot express $j$ as a countable set, and the sum has to be separated into a sum over the point spectrum, and an integration over the continuous spectrum. 

\subsection{Forward v.~backward, Schr\"odinger v.~Heisenberg}

The juxtaposition of Liouville (more generally, Perron--Frobenius) v.~Koopman representations can be understood as the forward picture of a dynamical system v.~the backward (adjoint) description of their observables. For simplicity, we first consider $\delta$-distributed initial conditions in this section. The solution of $g\l( x(t;x_0) \r)$, that is, the $g$ function evaluated on the trajectory of the dynamical system with an initial condition $x_0$ can be obtained by solving the partial differential equation \cite{chorin2002optimal}:
\begin{align}
    \frac{\partial}{\partial t} \gamma(x,t) ={}& \mathcal{L} \gamma(x,t) = \sum_{j=1}^N F_j \frac{\partial}{\partial x_j}  \gamma(x,t), \nonumber \\ \gamma(x,0):={}& g(x). \label{eq:adjLiouville}
\end{align}
Note that the function $g$ is evaluated at $t=0$ as the initial data. The solution $\gamma(x,t)$ characterizes how the function $g$ ``rides on'' the dynamics: it is elementary to show that $\gamma(x_0,t)=g(x(t;x_0))=\l(\mathcal{K}_t g\r) \l(x_0\r)$ is a spatiotemporal field of the $g$ function evaluated at the trajectory originated from $x_0$. The solution of $\gamma$ depends on $g$; by setting an identity function $g:=I$, i.e., $g(x)=x$, $\gamma$ is $\varphi$ in Eq.~\eqref{eq:formalSol}. Note that the above evolutionary equation \eqref{eq:adjLiouville} for $\gamma$ is also linear, again with the trade-off that one has to solve the infinite-dimensional \emph{field} $\gamma: \mathbb{R}^N \times \mathbb{R} \rightarrow \mathbb{R}$.

The duality between the forward Liouville/Perron--Frobenius description and the backward (adjoint) Koopman description is most clear when we consider the expectation value of $g$ with respect to the probability density $\rho\l(x, t\r)$ \cite{gerlich1973VerallgemeinerteLiouvilleGleichung}:
\begin{align}
     \int_{\mathbb{R}^N} g(x) \, \rho(x,t) \,\dd x ={}& \int_{\mathbb{R}^N} \gamma(x,0) \, \rho(x,t) \,\dd x  \nonumber \\ ={}& \int_{\mathbb{R}^N} \gamma\l(x, t\r) \, \rho\l(x,0\r) \, \dd x
\end{align}
Here, the first line is the forward picture where the density is evolving, and the second line the Koopman picture where the observables are evolving. Moreover, the adjoint nature of these two pictures can be transparently observed if we use the Dirac bra-ket notation in the Hilbert space $L^2\l(\mathbb{R}^N, \dd x\r)$:
\begin{subequations}
\begin{align}
    \frac{\dd}{\dd t} \l\langle g, 1\r\rangle_{\rho_t} ={}& \int_{\mathbb{R}^N} g(x) \, \mathcal{L}^\dagger \rho(x,t) \dd x \\
    ={}&\l\langle g, \mathcal{L}^\dagger \rho\l(t\r) \r\rangle
    = \l\langle \mathcal{L} \gamma\l(t\r) \large, \rho_0 \r\rangle\\
    ={}& \int_{\mathbb{R}^N} \mathcal{L} \gamma\l(x,t\r) \rho_0\l(x\r) \dd x  
\end{align}
\end{subequations}
The operator $\mathcal{L}$ is exactly the adjoint of the Liouville operator, $\mathcal{L}^\dagger$, in Eq.~\eqref{eq:Liouville}. In terms of the Koopman representation, $\mathcal{L}$ is exactly the infinitesimal Koopman operator $\mathcal{K} := \lim_{t\downarrow 0} \l(\mathcal{K}_t - 1\r)/t$ \cite{mauroy2016GlobalStabilityAnalysisa}. In terms of stochastic processes, the adjoint operator is the Kolmogorov backward operator, which is the infinitesimal generator of a stochastic process.

Beside general dynamical system theory and stochastic processes, the best-known exhibition of such duality are the two pictures in the theory of quantum mechanics. The forward/Liouville/Perron--Frobenius representation is analogous to the Schr\"odinger picture, in which the quantum state evolves forward in time. The backward/adjoint/Koopman picture is analogous to the Heisenberg picture, in which the operators (observables) dynamically evolve forward in time. 

\subsection{Koopman von Neumann mechanics}

Koopman von Neumann mechanics, hereinafter referred to as KvN mechanics, is a classical wave mechanics theory. Initially developed for Hamiltonian systems, this wave mechanics theory aims to bridge and unify the theories of classical and quantum mechanics. In most of the literature, Koopman and von Neumann were cited as the origin of KvN. However, we will see below that KvN is distinct from the backward/adjoint/Koopman representation in \cite{Koopman315,Koopman255,neumann1932ZurOperatorenmethodeKlassischen}, and is in fact a forward Perron--Frobenius-Schr\"odinger representation. 

Because classical Hamiltonian systems are the best and most intuitive way to understand KvN mechanics, we dedicate the next subsection only to Hamiltonian systems, with a following subsection generalizing the results to non-Hamiltonian systems.

\subsubsection{Special case: Hamiltonian systems}
The wavefunction is the fundamental description of a system's state in quantum mechanics. With the motivation of bridging the classical and quantum mechanics, instead of the central object $\rho(x,t)$ in the Liouville representation, KvN mechanics focuses on a complex-valued \emph{classical wavefunction} $\psi(x,t)$ with the Born rule $\rho(x,t)=\l\vert \psi (x,t) \r\vert^2 := \psi^\ast(x,t) \psi(x,t)$. Throughout this documentation, we reserve $\psi(x,t)$ to denote this classical KvN wavefunction and $\psi^\ast$ its complex conjugate. 

Consider a Hamiltonian system, whose state consists of $N$ pairs of generalized coordinates $q_j$ and momenta $p_j$, $j=1 \ldots N$. In previous notation $x=(q,p)$. A classical Hamiltonian ${H}(q,p):\mathbb{R}^N \times \mathbb{R}^N \rightarrow \mathbb{R}$ is defined and the system evolves according to Hamilton's equation:
\begin{subequations}
\begin{align}
\label{eq:ham1}
\dot{q}_j={}& \frac{\partial {H}\l(q,p\r)}{\partial p_j}, \\ 
\label{eq:ham2}
\dot{p}_j={}& -\frac{\partial {H}\l(q,p\r)}{\partial q_j},    
\end{align}
\end{subequations}
$j=1\ldots N$. The Liouville equation \eqref{eq:Liouville} of the system is thus
\begin{align}
    \frac{\partial}{\partial t}\rho(q,p,t) ={}& \sum_{j=1}^N \frac{\partial}{\partial p_j} \l[\frac{\partial {H}\l(q,p\r)}{\partial q_j} \rho \l(q,p,t\r) \r] \nonumber \\
    {}& - \sum_{j=1}^N \frac{\partial}{\partial q_j} \l[\frac{\partial {H}\l(q,p\r)}{\partial p_j}  \rho \l(q,p,t\r) \r]  \nonumber \\
    ={}& \sum_{j=1}^N \l(\frac{\partial {H}}{\partial q_j} \frac{\partial \rho}{\partial p_j} - \frac{\partial {H}}{\partial p_j} \frac{\partial \rho}{\partial q_j}\r). \label{eq:LiouvilleH}
\end{align}
In the last equation, we note that the special symplectic structure of the Hamiltonian led to the cancellation of the terms, $ \partial^2 H/ \partial q_j \partial p_j - \partial^2 H/ \partial p_j \partial q_j =0 $. 

The fundamental object in KvN mechanics is the KvN wavefunction $\psi: \mathbb{R}^N\times \mathbb{R}^N \times \mathbb{R} \rightarrow \mathbb{C}$ which maps $(x,p,t)$ to a complex-valued ``amplitude''. Immediately, one can show that the solution to a ``classical'' Schr\"odinger equation:
\begin{equation}
    \frac{\partial \psi(q,p,t)}{\partial t} = \sum_{j=1}^N \l(\frac{\partial {H}}{\partial q_j} \frac{\partial \psi}{\partial p_j} -\frac{\partial {H}}{\partial p_j} \frac{\partial \psi}{\partial q_j}\r), \label{eq:SchroedingerH}
\end{equation}
leads to the solution to the Liouville equation \eqref{eq:LiouvilleH}, with $\rho := \psi^\ast \psi$. A more formal derivation of Eq.~\eqref{eq:SchroedingerH} from operator axioms can be found in \cite{wilczek2015NotesKoopmanNeumann}.
One defines the generator of the dynamics of the wavefunction:
\begin{equation}
    \mathcal{L}^\dagger_\text{KvN} := - \sum_{j=1}^N \l(\dot{q}_j \frac{\partial}{\partial q_j} + \dot{p}_j \frac{\partial}{\partial p_j} \r),
\end{equation}
noting that it is identical to the generator of the dynamics of $\rho$, i.e., $\mathcal{L}$ in Eqs.~\eqref{eq:Liouville} and \eqref{eq:adjLiouville}. A KvN Hamiltonian is defined as 
\begin{equation}
    \mathcal{H}_\text{KvN} := i \mathcal{L}^\dagger_\text{KvN} = - i  \sum_{j=1}^N\l( \dot{q}_j \frac{\partial}{\partial q_j} + \dot{p}_j \frac{\partial}{\partial p_j} \r) \label{eq:generatorHamil}
\end{equation}
where $\dot q$ and $\dot p$ are given by Eqs.(\ref{eq:ham1},\ref{eq:ham2}) so that the evolutionary equation \eqref{eq:SchroedingerH} for the KvN wavefunction resembles the Schr\"odinger equation:
\begin{equation}
    i \frac{\partial}{\partial t} \psi(q,p,t) = \mathcal{H}_\text{KvN} \psi(q,p,t). \label{eq:Schroedingeri}
\end{equation}
Note that neither $\hbar$ nor a phase  is involved in this purely classical theory. The imaginary number $i$ is introduced, simply to make $H_\text{KvN}$ a Hermitian operator; see below. Equations \eqref{eq:SchroedingerH} and \eqref{eq:Schroedingeri} are completely equivalent, and KvN mechanics is \emph{fully classical}. Contrary to quantum mechanics, the 
phase of the Koopman wavefunction is irrelevant, and thus the theory cannot be used to explain the double-slit experiment or Aharonov--Bohm effect.

Given any two test wavefunctions $\psi_1(x,t)=\psi_1(q,p,t)$ and $\psi_2(x,t)=\psi_2(q,p,t)$ in $L^2(\mathbb{R}^{2N}, \dd x)$, using the Dirac bra-ket notation:
\begin{align}
    {}& \l \langle \mathcal{H}_\text{KvN}^\dagger \psi_1 (t)\r \vert \l. \vphantom{\mathcal{H}_\text{KvN}} \psi_2 (t) \r\rangle = \l \langle \vphantom{\mathcal{H}_\text{KvN}}  \psi_1 (t)\r \vert \l.\mathcal{H}_\text{KvN} \psi_2 (t) \r\rangle \nonumber \\
    ={}& \int_{\mathbb{R}^{2N}} \psi_1^\ast (x,t) \l[i \sum_{j=1}^N \l(\frac{\partial H}{\partial q_j} \frac{\partial }{\partial p_j} - \frac{\partial H}{\partial p_j} \frac{\partial }{\partial q_j} \r)\r] \psi_2(x,t) \, \dd x \nonumber \\
    ={}& \int_{\mathbb{R}^{2N}} \psi_2(x,t) \l[i \sum_{j=1}^N \frac{\partial H}{\partial q_j} \frac{\partial}{\partial p_j} - \frac{\partial H}{\partial p_j} \frac{\partial }{\partial q_j} \r]^\ast \psi_1(x,t) \, \dd x \nonumber \\
    ={}&  \l \langle \mathcal{H}_\text{KvN}  \psi_1 (t)\r \vert \l.\vphantom{\mathcal{H}_\text{KvN}^\dagger }\psi_2 (t) \r\rangle,
\end{align}
we thus conclude that $\mathcal{H}_\text{KvN}=\mathcal{H}_\text{KvN}^\dagger$, i.e., it is a Hermitian operator. Note that the negative sign of the complex conjugate of $i$ is cancelled by the negative sign from the integration by parts. We also assume the boundary terms are zero, imposed by the typical requirement of wavefunctions $\l\vert \psi\r\vert \rightarrow 0$ at the boundaries. The Hermitian property results in the unitary propagator $\exp\l(-i t \mathcal{H}_\text{KvN} \r)$, and the normalization $\int_{\mathbb{R}^{2N}} \psi^\ast(x,t)\, \psi(x,t)\, \dd x$ is preserved under the dynamics. Note that a Hermitian $\mathcal{H}_\text{KvN}$ implies an anti-self-adjoint $\mathcal{L}^\dagger_\text{KvN} = -i \mathcal{H}_\text{KvN}$. 

We emphasize that the expression resembles the formulation in quantum mechanics only at a superficial level: the resulting propagator is exactly the classical propagator $\exp\l(t \mathcal{L}_\text{KvN}\r)$, once we plug in $\mathcal{H}_\text{KvN} := i \mathcal{L}^\dagger_\text{KvN}$. The normalization is conserved because the Liouville operator $\mathcal{L}^\dagger_\text{KvN}$ only moves the probability mass in the phase space from one configuration to another. Note also that the generator $\mathcal{H}_\text{KvN}$ is not the operator form of the Hamiltonian in quantum mechanics which substitutes the momentum $p$ by $\l({\hbar}/{i}\r)\partial_q$: In the classical KvN mechanics, the momentum and position are commuting ($[{q}_j, {p}_j]=0$), in contrast to the  canonical commutation relation between the position and momentum operators in quantum mechanics, $[\hat{q}_j, \hat{p}_j]=i\hbar \hat{I}$ (we use $\hat{\cdot} $ to denote quantum-mechanical operators). 

\subsubsection{General nonlinear dynamical systems}

With the same spirit, for a general dynamical system following \eqref{eq:Liouville}, it is possible to prescribe a classical Schr\"{o}dinger equation \cite{bogdanov2014StudyClassicalDynamical,bogdanov2019QuantumApproachDynamical,joseph2020KoopmanNeumannApproach}:
\begin{equation}
    \frac{\partial \psi(x,t)}{\partial t} = - \sum_{j=1}^N \l[F_j(x) \frac{\partial }{\partial x_j} + \frac{1}{2} \frac{\partial F_j(x)}{\partial x_j}\r] \psi\l(x,t\r), \label{eq:SchroedingerF}
\end{equation}
such that $\rho(x,t) := \psi^\ast(x,t) \psi(x,t)$ solves \eqref{eq:Liouville}. Note that the state space is $\mathbb{R}^N$, so the KvN wavefunction $\psi:\mathbb{R}^N \times \mathbb{R} \rightarrow \mathbb{C}$ (in contrast to the $2N$-dimensional Hamiltonian systems). Because the KvN Hamiltonian, defined as
\begin{equation}
\mathcal{H}_\text{KvN} := 
-i \sum_{j=1}^N \l[ F_j(x) \frac{\partial }{\partial x_j} + \frac{1}{2} \frac{\partial F_j(x)}{\partial x_j}\r] , \label{eq:generatorF0}
\end{equation}
is Hermitian, the propagator is again unitary and the normalization is preserved under the dynamics. 
It is tempting to define $\partial_{x_j}$ in \eqref{eq:SchroedingerH} or \eqref{eq:SchroedingerF} as auxiliary operators $\mathcal{P}_j:=-i\partial_{x_j}$ \cite{bogdanov2014StudyClassicalDynamical,bogdanov2019QuantumApproachDynamical,joseph2020KoopmanNeumannApproach} and symmetrize the operator into the form 
\begin{equation}
    \mathcal{H}_\text{KvN}=\frac{1}{2}\sum_{j=1}^N \l(\mathcal{P}_j F_j(x) + F_j(x) \mathcal{P}_j \r). \label{eq:generatorF}
\end{equation} 
Here, the $\mathcal{P}_j$ operates on everything to its right, i.e., for two test functions $f$ and $g$, $\mathcal{P}_j \l(fg\r):=\l(\mathcal{P}_j f\r)g + f \l(\mathcal{P}_j g\r)$.
Note that \emph{this auxiliary operator is not the momentum of a Hamiltonian system}, that is, $p_j$ in \eqref{eq:SchroedingerH}: the position $x_j$ commutes with $p_j$ in KvN mechanics, but not with $\mathcal{P}_j$. It is most clear in the axiomatic construction of the KvN mechanics in \cite{wilczek2015NotesKoopmanNeumann}, in which $\partial_{x_j}$ is denoted by an operator $\lambda_{x_j}$ and is not associated with the classical momenta. Identifying $\partial_{x_j}$ as a momentum, such as in \cite{joseph2020KoopmanNeumannApproach}, should not be considered as a fully classical KvN mechanics. Such an indentification would invoke the quantum mechanical Stone-von Neumann theorem, which specifies the momentum operator $\mathcal{P}_j=-i\partial_{x_j}$ (in the unit $\hbar=1$) in the position representation and results in the Heisenberg uncertainty principle. In terms of KvN mechanics, $\mathcal{P}_j$ should be considered as merely an operator in the Hilbert space. 

For completeness, we present a further, alternative construction of KvN mechanics that we have not seen in the quantum literature, that uses the fact that one can construct a Hamiltonian system in an augmented space $\mathbb{R}^{2N}$ for any non-Hamiltonian system \eqref{eq:ODE}. To achieve this, we embed the $N$-dimensional $x$-space into a $2N$-dimensional $(x,\varpi)$-space, where $\varpi$ will be the classical momenta that are associated with $x$ in the auxiliary Hamiltonian system. Next, we prescribe an auxiliary Hamiltonian
\begin{align}
    H_\text{aug}(x,\varpi):= \sum_{j=1}^N \varpi_j F_j=\frac{1}{2}\sum_{j=1}^N \l(\varpi_j F_j + F_j \varpi_j\r). \label{eq:Haug}
\end{align}
Despite Eq.~\eqref{eq:Haug} resemblance to the form of Eq.~\eqref{eq:generatorF}, it is important to realize that they are not the same: $\mathcal{P}_j:=-i\partial_{x_j}$ and $x$ are non-commuting, but the classical $\varpi_j$ and $x_j$ are commuting. 
Hamilton's equation prescribes the evolution of $x$ and $\varpi$:
\begin{subequations}
\begin{align}
    \dot{x}_j ={}& \partial_{\varpi_j} \mathcal{H}_\text{aug} = F_j\l(x\r),  \\
    \dot{\varpi}_j ={}& - \partial_{x_j} \mathcal{H}_\text{aug} = -\sum_{k=1}^N \varpi_k \frac{\partial F_k \l(x\r)}{\partial x_j} . 
\end{align}
\end{subequations}
Note that the solution of $x$ is consistent with that of \eqref{eq:ODE}. 
However, following the  KvN construction on this classical Hamiltonian system leads to a distinct generator from Eq.~\eqref{eq:generatorF}:
\begin{equation}
    \mathcal{H}_\text{KvN}^{(2N)} = - \frac{i}{2} \sum_{j=1}^N \l[F_j\l(x\r) \frac{\partial}{\partial x_j} - \sum_{k=1}^N \varpi_k \frac{\partial F_k \l(x\r)}{\partial x_j} \frac{\partial }{\partial \varpi_j}\r], \label{eq:generatorF2N}
\end{equation}
with the Schr\"odinger equation
\begin{equation}
    i\frac{\partial}{\partial t} \psi\l(x, \varpi, t\r) =   \mathcal{H}_\text{KvN}^{(2N)} \psi\l(x, \varpi, t\r)
\end{equation}
Note that the KvN wavefunction $\psi(x,\varpi,t)$ in this construct lives in a higher-dimensional space, $\mathbb{R}^{2N}$; we thus decorate the Hamiltonian by a superscript $(2N)$. The auxiliary operators are now $-i\partial_{x_j}$ and $-i\partial_{\varpi_j}$, and the classical conjugate momentum is $\varpi$.

In the discussion below, we will strictly focus on $\mathcal{H}_\text{KvN}$ which lives in a smaller dimension and leaves the existence of $\mathcal{H}_\text{KvN}^{(2N)}$ as an intriguing observation. 

\subsubsection{Spectral analysis of the KvN Hamiltonian}
It is natural to solve the linear KvN Schr\"odinger equation by identifying the eigenvalue $\nu_j$ and eigenvector $\chi_j$ pairs satisfying
\begin{equation}
    \mathcal{H}_\text{KvN} \chi_j = \nu_j \chi_j. 
\end{equation}
Because $\mathcal{H}_\text{KvN}$ is Hermitian, the eigenvalues $\nu_j$'s are real and the eigenvectors are mutually orthogonal $\l\langle \chi_j, \chi_j\r\rangle=\delta_{i,j}$. Again, the Hilbert space here is $L^2\l(\mathbb{R}^N, \dd x\r)$. The full solution can then be expressed as
\begin{equation}
    \psi(x,t) = \sum_{i} \kappa_j e^{-i \nu_j t } \chi_j(x),
\end{equation}
where $\kappa_j$'s are the complex-valued coefficients satisfying the initial condition, $\kappa_j = \l\langle \chi_j, \psi(t=0) \r\rangle$.

If (\ref{eq:ODE}) is linear, then the spectrum of $x$ is related to the spectrum of $\rho(x,t)$ \cite{rowley2009spectral}. For KvN, the relationship between spectra is not clear; note that the transformation from $\rho(x,t)$ to $\psi(x,t)$ is nonlinear (quadratic). More discussion of the spectral differences is given in the next section.

\subsection{Differences between the Koopman representation and Koopman von Neumann mechanics}

We are now in a position to list some of the major differences between the Koopman representation and KvN mechanics. Our central result is that Koopman von Neumann mechanics is not a Koopman representation. The most fundamental observation which sets the tone of this statement is: The \emph{Koopman representation is the Heisenberg/adjoint/backward picture, and Koopman von Neumann mechanics is the Schr\"{o}dinger/forward picture.} Despite their being two sides of the same coin, the two interpretations create ramifications below.

\noindent{\bf Different objects evolving in time.}
In the Koopman representation, we aim to describe the evolution of observables, which are integrable functions of an initial condition with respect to an initial distribution; in Koopman von Neumann mechanics, we aim to describe the evolution of a KvN wavefunction, which is related to a probability distribution. Note that the symbols $\psi$ (and $\varphi$) in the original Koopman paper \cite{Koopman315} are functions of the initial microscopic state, and not a wavefunction, which is a function of state space \emph{and} time. 

\noindent{\bf Different inner-product spaces.}
The Koopman representation lives in the Hilbert space $L^2 \l(\mathbb{R}^N, \rho_0 \l(x\r) \dd x\r)$, equipped with the inner product $\l\langle\cdot, \cdot \r\rangle_{\rho_0}$, that is, Eq.~\eqref{eq:innerProduct} with $\dd \mu= \rho_0 \l(x\r) \dd x$. For ergodic systems, it is common to initiate the system at the invariant measure and define the inner product by the invariant measure $\dd \mu= \rho_\text{inv} \l(x\r) \dd x$. Note that in \cite{Koopman315}, the probability density $\rho$ is always there in the inner product. In contrast, the KvN mechanics lives in the Hilbert space $L^2 \l(\mathbb{R}^N, \dd x\r)$ equipped with the inner product $\l\langle \cdot, \cdot \r\rangle$ between two KvN wavefunctions, that is, Eq.~\eqref{eq:innerProduct} with the Lebesgue measure $\dd \mu=  \dd x$.

\noindent{\bf Different generators and propagators.} 
Koopman \cite{Koopman315} showed that for Hamiltonian systems, the propagator for the observables, the finite-time Koopman operator $\mathcal{K}_t$ is unitary (and thus Koopman denoted it by $U_t$), and deduced via the Stone's theorem on one-parameter unitary groups that the infinitesimal Koopman operator $\mathcal{K} := \lim_{t\downarrow 0} \l(\mathcal{K}_t - 1\r)/t$ is anti-self-adjoint (in the original paper, $\mathcal{K} =iP$ is anti-self-adjoint because $P$ is self-adjoint/Hermitian). For general irreversible systems, it is known that the Koopman operator does not have to be unitary (see examples in \cite{williamsDataDrivenApproximation2015}). As for KvN mechanics, the generator of the KvN wavefunction dynamics (i.e., $-i\mathcal{H}_\text{KvN}$) is always anti-self-adjoint, regardless of whether $F$ is a Hamiltonian flow. In the case of a Hamiltonian system, the divergence term $\sum_{j=1}^N \partial_{x_j} F_j = 0$ because the flow of the full state $x=(q,p)$ is volume conserving, and Eq.~\eqref{eq:generatorF0} is identical to Eq.~\eqref{eq:generatorHamil}.

\noindent{\bf Different spectra and basis functions.} For a general irreversible system, the Koopman operator $\mathcal{K}_t$ does not have to be unitary, so $\vert e^{\tau \lambda_j} \vert $ are not necessarily $1$ for any $\tau \ge 0$. That is, observables do not conserve their norm under the dynamics. Koopman eigenfunctions $\phi_j$'s are not necessarily orthogonal to each other. Note that the Koopman spectrum is shared with its adjoint, the Perron--Frobenius transfer
operator and the forward infinitesimal operator $\mathcal{L}^\dagger$ \cite{klus2018DataDrivenModelReduction}. In terms of KvN mechanics, by construction $\mathcal{H}_\text{KvN}$ is Hermitian, which guarantees $\vert e^{i \tau \nu_j} \vert=1$ for any $\tau \ge 0$ and orthogonal $\chi_j$'s. That is, observables conserve their norm under the dynamics. The KvN decomposition is closely related to the Proper Orthogonal Decomposition (POD) \cite{lumley2007stochastic,sirovich1987turbulence,berkooz1993ProperOrthogonalDecomposition}. As concisely summarized in Schmid's original paper \cite{schmid2010DynamicModeDecomposition} on Dynamic Mode Decomposition, which is a Koopman representation but restricted only to linear observables, the Koopman representation attempts to capture the orthogonality ``in time'', and the POD (and the KvN mechanics here) attempts to capture the orthogonality ``in space''. 

\noindent{\bf KvN mechanics is a wave mechanics with perpetually oscillating eigenfunctions, which may not be true for Koopman eigenfunctions.} A natural and interesting consequence of the differences of the spectra of these two formulations is that, the solution in KvN mechanics is purely composed of oscillatory eigenfunctions, while the Koopman eigenfunctions can be both decaying and oscillatory. Let us consider a simple, one-dimensional decaying dynamics, $\dot{x}=-x^2$ on a strictly positive domain $x\in\mathbb{R}^+$, which was presented as an illustrative example in \cite{lin2021data}. We can show that a Koopman eigenfunction of this problem is $g(x)=\exp\l(-1/x\r)$ and $\mathcal{K} g = -g $, and $\lambda=-1$ clearly indicating the dissipating nature of the dynamics. On the other hand, the KvN wavefunctions are always oscillatory. Similar to using Fourier modes to describe exponentially decaying signals, in the $\dot{x}=-x^2$ example, to describe the dynamics of a single Koopman eigenfunction $g(x)=\exp\l(-1/x\r)$ requires infinitely many KvN eigenfunctions. 

\noindent{\bf Relationship between two operators $\mathcal{H}_\text{KvN}$ and $\mathcal{K}$.} We can use the evolutionary equation of the expectation value of an observable $g :\mathbb{R}^N\rightarrow \mathbb{R}$ with respect to the probability density $\psi^\ast(x,t) \psi(x,t)$ to establish the formal relationship between the infinitesimal Koopman operator $\mathcal{K}$ and the KvN Hamiltonian $\mathcal{H}_\text{KvN}$ (below, integrals are over the entire space $\mathbb{R}^N$):
\begin{equation}
    \frac{\dd}{\dd t}\l\langle g \r\rangle(t)  = \frac{\dd}{\dd t} \int \psi^\ast\l(x,t\r) g\l(x\r)\, \psi\l(x,t\r) \dd x.
\end{equation}
Using the KvN Schr\"odinger equation,
\begin{align}
    \frac{\dd}{\dd t}\l\langle g \r\rangle(t) ={}& \int\l(- i \mathcal{H}_\text{KvN} \psi \r)^\ast g \, \psi \, \dd x \nonumber \\ {}& + \int \psi^\ast \, g \l(-i \mathcal{H}_\text{KvN} \psi\r) \dd x.
\end{align}
We next use the self-adjointness of the operator $\mathcal{H}_\text{KvN}$ to transfer the $\mathcal{H}_\text{KvN}$ in the first term to the observable $g$:
\begin{align}
    \frac{\dd}{\dd t}\l\langle g \r\rangle{}&(t) = i \int \psi \l( \mathcal{H}_\text{KvN} g - g \mathcal{H}_\text{KvN}\r) \psi\, \dd x.
\end{align}
We now switch to the adjoint/backward/Heisenberg picture to make the connection to the Koopman representation. The observable is now time-dependent, i.e., $\gamma(x,t)$ in Eq.~\eqref{eq:adjLiouville}, such that
\begin{equation}
    \frac{\partial}{\partial t} \gamma \l(x,t\r) = i \l[\mathcal{H}_\text{KvN} \gamma \l(x,t\r)- \gamma \l(x,t\r) \mathcal{H}_\text{KvN}\r].
\end{equation}
Next, plugging in the definition of $\mathcal{H}_\text{KvN}$, Eq.~\eqref{eq:generatorF0}, on a test function $\psi$:
\begin{align}
    \mathcal{H}_\text{KvN} \gamma \psi ={}& - i \l[ \sum_{j=1}^N F_j \frac{\partial \gamma}{\partial x_j} \psi + F_j \gamma \frac{\partial  \psi}{\partial  x} + \frac{1}{2} \frac{\partial F_j}{\partial x_j} \r], \nonumber \\
    \gamma \mathcal{H}_\text{KvN} \psi ={}& - i\gamma\l[ \sum_{j=1}^N F_j \frac{\partial \psi}{\partial x} + \frac{1}{2} \frac{\partial F_j}{\partial x_j} \r],
\end{align}
noting that $F_j(x) \gamma(x,t) =  \gamma(x,t) F_j(x)$, we arrive at the operator equation
\begin{align}
    \frac{\partial}{\partial t} \gamma\l(x,t\r) ={}& \sum_{j=1}^N F_j(x) \frac{\partial \gamma(x,t)} {\partial x_j} \nonumber \\
    ={}& \l[\mathcal{L} \gamma\r] \l(x,t\r) = \l[\mathcal{K} \gamma\r] \l(x,t\r).
\end{align}
Finally, we identify the relationship between the infinitesimal Koopman operator and the KvN Hamiltonian: given any test observable $\gamma=\gamma\l(x,t\r)$, 
\begin{equation}
 \mathcal{K} \gamma = i\l(\mathcal{H}_\text{KvN}\gamma- \gamma\mathcal{H}_\text{KvN}\r). \label{eq:H-L}
\end{equation}
The fact that the right hand side of the equation resembles the Heisenberg equation should not be a surprise: the infinitesimal Koopman operator \emph{is} the generator of the operator dynamics. 

\section{Quantum computation for solving dynamical systems}\label{sec:quantumComputation}

There are (at least) two difficulties when solving nonlinear equations using a classical computer: (i) The original system of interest is very high dimensional, such that classical computers run into memory issues, but perhaps the system can be encoded in the exponentially sized Hilbert space of a quantum computer. (ii) The system of interest is not necessarily high dimensional but is highly nonlinear (maybe chaotic), such that classical computers run into precision issues.

We have already established that finite dimensional nonlinear systems can be mapped to infinite dimensional linear systems which can then be approximated in a reduced computational space by finite dimensional linear systems using the methods outlined in the previous section. 
Since the nonlinearity is absent in this approach, it might appear that precision is not a problem and we don't need to worry about (ii). This is not the case, however, because the linear dynamics of the reduced system might not represent the nonlinear dynamics faithfully. On the other hand, one might hope that this issue can be alleviated by using very large dimensional linear systems. 
Quantum computation does have the advantage that the dimension of the operational space scales exponentially with the number of underlying computational units (e.g., qubits); thus, the hope is that a sufficiently large space could allow us to overcome both difficulities (i) and (ii).

One tempting classical way to numerically solve an initial-value problem for a complex nonlinear dynamical system is to integrate the evolutionary equation \eqref{eq:ODE} forward in time in the phase-space picture. Consider explicit Euler time-integration, with a uniform time step $\Delta$. The numerical process entails: (1) start from the initial state, $x_0$, (2) evaluate the RHS of Eq.~\eqref{eq:ODE} at time $0$, (3) compute the state at time $\Delta$ by $x_\Delta:=x_0 + F\l(x_0\r) \Delta$, (4) iterate steps (2-3) until the terminal time $T$. 

Importantly, for nonlinear systems, step (2) requires a nonlinear operation on $x$ to compute the flow at subsequent times, $F(x_{m\Delta})$. For this scheme, we cannot bypass the nonlinear operation. 

When we implement this procedure on a quantum computer an exponential number of initial states (in the number of time steps) are required to integrate the equation \cite{leyton2008quantum}. 
This can be seen in the simple case of a quadratic nonlinearity, where the variable $x$ is encoded in a quantum state $|\psi_t \rangle=\sum_i\psi_{t,i} |i\rangle $ via $x_i(t) \propto \psi_{t,i}$. 
Due to linearity, two copies of the state $| \psi_{t} \rangle$ at time $t = m\Delta$ are needed to prepare a quantum state with amplitudes proportional to $x_i(t)^2$. 
Moreover, the algorithm that achieves this consumes the two copies in the process~\cite{leyton2008quantum}. 
On the other hand, copies can not be created on demand due to the no-cloning theorem: they have to be prepared individually using the same procedure instead.   
Thus, we see that, for an integration time $T = N\Delta$, of order $2^N$ copies of $| \psi_0 \rangle$ will be needed to integrate the equation. As such, this approach is not computationally feasible on a quantum computer.

Instead, the linearity of the representations presented in the previous section opens a door for using quantum computation (QC) for solving nonlinear dynamical systems. If the resulting linear differential equations, expressed as $\dot{\mathbf{x}}(t) = \mathbf{L} \cdot \mathbf{x}(t)$, satisfy some technical requirements (see \cite{berry2014HighorderQuantumAlgorithm} for details), we can leverage quantum algorithms for (1) simulating the finite-time propagator $e^{t \mathbf{L}}$ if $i \mathbf{L}$ is Hermitian \cite{berry2007EfficientQuantumAlgorithms} and (2) an algorithm proposed by Berry for solving general linear differential equations \cite{berry2014HighorderQuantumAlgorithm}, if $i \mathbf{L}$ does not have the desired Hamiltonian structure. KvN mechanics produces a set of equations that can be solved with both (1) and (2), but the Koopman representation will in general require the more general approach (2) unless the Koopman operator, itself, is Hermitian.

Potential pitfalls and technical difficulties exist even with perfect QC implementations for solving linear dynamical systems. 
This is mainly due to a trade-off between linearity and the dimension of the representation.
In order to gain linearity in these representations, the dimensionality of the operational space is generally infinite. In practice, it is necessary to perform some form of truncation, projection, or discretization, which reduces the computational space to finite-dimensionality but inevitably induces errors. 

We will use the toy example $x(t)=-x^2(t)$ in each of the representations. We provide parallel analyses on an alternative model, the two-dimensional Van der Pol oscillator, in Appendix \ref{app:2}, showing our observations below are generic and not dependent on the specific toy example we chose. We believe that the conclusions we draw from the examples will be applicable to high-dimensional systems as well.

\subsection{Encoding}
\black{
It is useful to discuss different means of quantum encoding for different linear representations presented in previous sections. In quantum computing, we encode information by the amplitude of a set of bases of a quantum state. More precisely, suppose we use the Dirac bra-ket notation to express a quantum state $\l \vert \psi_Q \r\rangle $ by a set of bases $\l \vert i_Q \r\rangle$:
\begin{equation}
    \l\vert \psi_Q \r\rangle= \sum_{i} a_i \l \vert i_Q \r\rangle. \label{eq:encoding}
\end{equation}
We use the subscript $Q$ to differentiate the quantum states from classical KvN wavefunctions. Then, the amplitude $\left\{a_i\right\}_i$ will be used to encode the variables that we would like to evolve in a dynamical system. 

There is a natural choice of variables to evolve for each of the linear representations. For the Liouville representation, the fundamental dynamic variable is the probability density. As such, the most natural choice is to use $a_i$ to encode either the probability $p\left(s_i\right)$ (up to normalization) that the state is in a discrete set $s_i$ in a continuous state space\footnote{For example, the probability of a continuous state $x$ is in an interval $s_i:=(x_i- \Delta x/2, x_i + \Delta x/2]$ where $\Delta x$ is the lattice size.} $S$ (where $s_i$'s are non-intersecting and the union of these sets covers the state space, $\cup_i S_i \supseteq S$), or the probability that the system is in a state $s_i$ in a discrete state space.
For Koopman von Neumann mechanics, thanks to its almost identical formulation to quantum mechanics, it is most natural to just use $a_i$ to encode the amplitude $\psi(x_i,t)$ after discretizing the $x$-space (see the classical Schr\"odinger equation Eq.~\eqref{eq:SchroedingerF}). Such encoding is very similar to the typical position representation of quantum mechanics. As for the Koopman representation, it is most natural to use $a_i$ to encode the fundamental dynamic variables, an observable evaluated on a specific initial condition, i.e., $\left(\mathcal{K}_t g_i\right)(x_0) $.

Because the state space $\l\vert \psi_Q \r\rangle$ grows exponentially as we increase the number of qubits, quantum computation enjoys an exponential advantage, which translates to different benefits in different representations. For the forward Liouville and Koopman von Neumann representation, we will enjoy an exponentially resolved state partition/discretization, which would be desired for continuous-state systems. For the backward Koopman representation, we will enjoy an exponentially larger set of observables, which is a desired property for the Koopman representation as its operational Hilbert space is formally infinite-dimensional. 

We remark that the above discussion on encodings is restricted to the ODE systems of the form Eq.~\ref{eq:ODE}, which is the primary focus of this manuscript. For those systems described by partial differential equations (PDEs), there exists an additional nuance of choosing either the Eulerian or Lagrangian specification to describe the system. 
}

\subsection{The Koopman representation}
The central idea of using the Koopman representation in QC is its linear evolution in the observable space. As such, given a complete set of observables, evaluated at the initial condition, the IVP is a linear problem which can, in principle, be solved with a quantum linear systems algorithm. However, there are two major challenges. First, observables can contain nonlinear functions of the initial condition, thus to enable end-to-end QC, one needs a general circuit for arbitrary nonlinear transformation of the initial condition $x$ to $g(x)$. This is a challenging problem, and will, necessarily, require multiple copies of an initial state, however, it does not necessarily lead to the same exponential blow-up in resources as the Leyton-Osborne approach \cite{leyton2008quantum}, as the generation of the nonlinear initial state (and possibly the inverse transformation at the end of the computation) only need be performed once \cite{holmes2023nonlinear}. Second, it is not necessarily true that the quantity of interest can be embedded in a finite set of observables, i.e., the minimal Koopman invariant space is not guaranteed to be finite dimensional. 

Even with the existence of a finite set, the set is not known \emph{a priori} and the number of observables spanning the invariant space may be very large. In any case, a certain approximation by a smaller set of observables is needed, which induces the closure problem. Although such a problem can be solved by data-driven projection when simulation data are available, in our case, the purpose of solving the IVP is to generate simulation data. As such, the projection operator of the closure scheme must be sought via other system-specific means. Furthermore, the approximation often has a short accurate predictive horizon. Higher-order methods such as time-embedded DMD or Mori--Zwanzig formalism \cite{lin2021data} exist, but they do not fully resolve the closure problem. 

\begin{figure}[!t]
\centering
    \includegraphics[width=0.8\columnwidth]{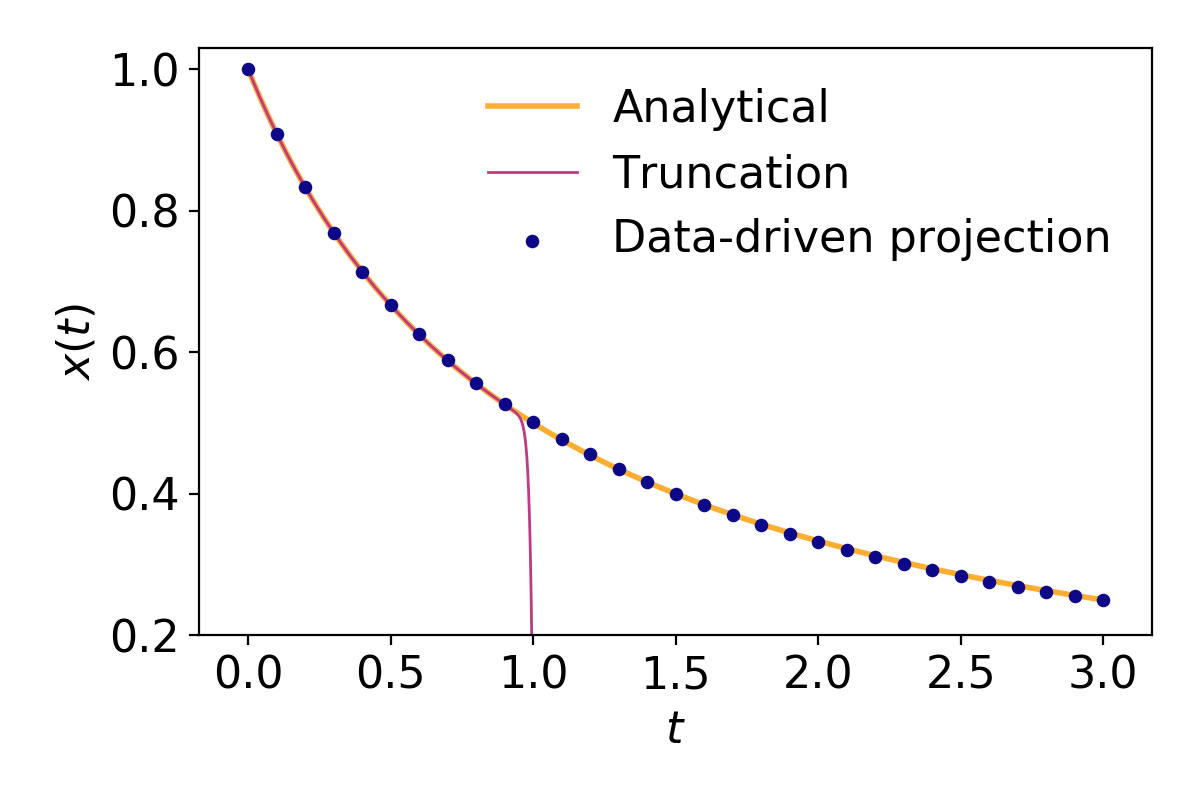}
    \caption{Carleman linearization with truncation at the first hundred order and approximate Koopman projection to a subfunctional space spanned by only the first three order.}
    \label{fig:carleman}
\end{figure}

\noindent{\emph{Example.}} Consider $\dot{x}=-x^2$, and suppose we choose the polynomial observables, $g_n(x):=x^n$, $n=0,1,\ldots.$ This is the Carleman linearization. The linear equation of the observables is $\dot{g}_n=-n g_{n+1}$, $n=1,\ldots$ 
When solving this problem with a QC, we encode the vector $g=(g_0,g_1,\dots)$ as a quantum state $|\psi(t) \rangle \propto \sum_n g_n \l(x(t)\r)  |n\rangle$. The goal of the algorithm is to output this state, which can then be used to compute properties of the solution, which is encoded in one of the amplitudes, $g_1\l(x(t)\r)$.~\footnote{More generally $x=(x_1,\ldots,x_N)\in \mathbb{R}^N$ as in \eqref{eq:ODE}, and the observables would take the form $x_1^{n_1}x_2^{n_2}\dots$ -- see \cite{liu2021efficient}. In that case the quantum algorithm prepares a quantum state whose amplitudes are proportional to all these observables. The solution can be obtained by projecting this state into a subsystem corresponding to $x=(x_1,\ldots,x_N)\in \mathbb{R}^N$ alone.}

Because $\dot{g}_n$ always depends on $g_{n+1}$, a proper closure scheme is needed. To showcase potential problems that can arise in closure schemes, we consider (1) truncating at $100^\text{th}$ order, i.e., $\dot{g}_{100}=0$ or equivalently $g_{101}=0$ and (2) projecting the infinite-dimensional series onto the first three polynomial functions, that is, assuming $\l[{\bf g}\l(x\l(t+\delta\r)\r)\r] \approx \mathbf{K}_\delta \cdot \l[{\bf g}\l(x\l(t\r)\r)\r]$ where $\l[g\r]$ is the stacked column vector of $g_0$, $g_1$, and $g_2$ and $\mathbf{K}_\delta$ is the approximate finite-time ($\delta$) Koopman operator. We remark that other closure schemes are possible; we do not aim to present an exhaustive analysis on closure schemes here, but we aim to use our examples to illustrate potential problems. We also note here that, many sophisticated moment closure schemes require nonlinear operations, e.g., $g_{101}\approx g_{50} g_{51}$. As explained above, it is not desirable to deploy nonlinear operations for integrating/evolving the system forward in time. Closure schemes using linear superpositions of the observables in the set to approximate the $g_{101}$, which is the essence of our second scheme leveraging projection via the EDMD procedure \cite{williamsDataDrivenApproximation2015} when data is available, outlined below. 

Assuming perfect snapshots of $\l[{\bf g}\l(x\l(t\r)\r)\r]$ are available at $t=0, \delta, 2\delta, \ldots $ for us to minimize the $L^2$ norm between the true solution and this projected dynamics in this 3-dimensional projected space, we use the minimizer $K_\delta$ for integrating the linear system forward for making prediction. Here, we choose $\delta=0.1$, and use the analytical solution to numerically solve for the minimizer $K_\delta$:
\begin{equation}
    \mathbf{K}_{\delta=0.1}=\begin{bmatrix}
  1. & -0. & 0.\\
  0.001 & 0.992 & -0.084\\
  -0.025 & 0.156 & 0.699\\
\end{bmatrix}.
\end{equation}
The aim of this analysis is to show the quality of prediction of a three-dimensional linear approximation of the infinite-dimensional Koopman representation, \emph{if we can obtain the approximate propagator $\mathbf{K}_{0.1}$ by means other than QC.} 

Both closure schemes lead to the standard linear problem of the form $\dot{\mathbf{x}} = \mathbf{L}\cdot \mathbf{x}$. Here, the elements of $\mathbf{x}$ are the $g_i$'s; $\mathbf{L}$ is a sparse, superdiagonal matrix in the first closure scheme, and simply $\mathbf{K}_\delta$ with the second closure scheme. In both cases, we can use Berry's algorithm or improvements thereof \cite{berry2014HighorderQuantumAlgorithm, berry2017quantum} to solve the linear problem. To showcase the potential problem arising from the closure schemes, we use the high-accuracy general-purposed integration \texttt{scipy.optimize.solve\_ivp} with \texttt{lsoda} to mimic a fault-tolerant QC implementation of the Berry's algorithm, in the limit of extremely small time steps. We stress that all of our numerical results in this study were generated on a classical computer, in order study the properties of the numerical solution in an idealized setting. We assume $x(0)=1.0$ and evolve the system to $t=3$. The result can be seen in Fig.~\ref{fig:carleman}, that this typical truncation scheme would be problematic for a long prediction horizon even with a very high order closure. However, the projection approach is more controllable even with a small (only three) set of observables considered. The apparent chicken-and-egg problem of using a QC and an approximate Koopman operator is that the approximate Koopman operator $\mathbf{K}_\delta$ of a complex dynamical system is generally not analytically tractable, so we need to rely on a data-driven approach to retrieve it numerically from the simulated trajectories. If we can already simulate the trajectories of a complex dynamical system, it is unclear what the advantage of a QC would be.

Note that the results in Fig.~\ref{fig:carleman} are consistent with the error bound derived in Ref. \cite[Eq. (27)]{forets_explicit_2017}, which for this problem, reduces to
\begin{equation}
    \| \varepsilon(t) \| \le \frac{t^n}{1-t}\,,
\end{equation}
which clearly is unbounded as $t\rightarrow 1$. A more detailed analysis of the conditions for finite-time blowup of Carleman are given in Appendix \ref{app:carleman}.

Finally, as pointed out in \cite{lin2021data}, the system has a one-dimensional Koopman invariant space $g_\ast(x)=\exp(-1/x)$, which follows $\dot{g}_\ast = - g_\ast$. Suppose we know such an invariant space (not necessarily one dimensional) and wish to leverage it to perform a classical simulation on a QC, we would need the nonlinear transformation $g_\ast$ to initiate the observable, linearly integrating it forward in time by QC, and then applying the nonlinear inverse transformation $g^{-1}_\ast(x)=-1.0/\log(x)$ to obtain $x$. 

It is currently challenging to implement both of these nonlinear transformations on a QC \cite{holmes2023nonlinear}. Such an invariant space idea is tightly connected to the idea of constant of integration methods of integrable systems. For example, it is known that for a nonlinear Burgers' equation, we can perform the Cole--Hopf transformation (c.f.~our $g_\ast$ for the $\dot{x}=-x^2$ model) to obtain a linear partial differential equation of the transformed variables (c.f.~our $\dot{g}_\ast = -g_\ast$ for the $\dot{x}=-x^2$ model). After the linear evolution in the transformed variable space, we can apply the inverse Cole--Hopf transformation (c.f.~$g^{-1}_\ast$ for the $\dot{x}=-x^2$ model) back to the physical space. 

The Cole-Hopf transform is a special case of a diagonalizable Koopman system. These systems are analogous to  {\it fast forwardable} systems in the quantum literature \cite{gu2021fast}, which have special time-energy uncertainty relations \cite{atia2017fast}. 

Clearly, the major bottleneck of this type of approach resides in the implementation of the nonlinear observables. This may be done either via direct computation \cite{verstraete2009quantum,gu2021fast}, or quantum machine learning methods may be used to attempt to find a diagonalization using optimization methods \cite{cirstoiu2020variational,commeau2020variational,gibbs2021long}. However, we note that the nonlinear observables need only be implemented once to map the initial state to the transformed space, then once more to invert the transformation. This is in contrast to the Leyton-Osborne method \cite{leyton2008quantum} that requires a number of nonlinear transforms that is proportional to the total evolution time.

\subsection{KvN mechanics}

\label{sec:KvN}

The commutator relation between the classical $x$ and $\mathcal{P}:=-i\partial_{x_i}$ operator, $\l[x, \mathcal{P}\r]=i$ resembles that between the position ($\hat{X}$) and momentum ($\hat{P}$) operators in quantum mechanics, $\l[\hat{X}, \hat{P}\r]=i$ (in the unit $\hbar=1$). Such an observation motivates the idea of realizing a quantum-mechanical Hamiltonian
\begin{equation}
    H_\text{QM}\l(\hat{P}, \hat{X}\r):= \frac{1}{2}\sum_{j=1}^n \l(\hat{P}_j F_j(\hat{X}) + F_j(\hat{X}) \hat{P}_j \r) \label{eq:generatorQ}
\end{equation}
for simulating a classical system on a QC.
Here, we use the notation $F_j(\hat{X})$ to denote a nonlinear map of the position operator which resembles the nonlinear flow, potentially through series expansions. For example, for the quadratic flow $F(x):=-x^2$, $F\l(\hat{X}\r)=\hat{X}^2$; for the sinusoidal flow $F(x)=\sin(x)$, $F\l(\hat{X}\r):= \hat{X}-\hat{X}^3/3!+\hat{X}^5/5!\ldots $ 
Then, the quantum mechanical wavefunction $\psi_\text{QM}$ of the quantum mechanical Schr\"odinger equation 
\begin{equation}
    i\frac{\partial}{\partial t} \psi_\text{QM}(x,t):= H_\text{QM}\l(\hat{P}, \hat{X}\r) \cdot \psi_\text{QM}\l(x,t\r) \label{eq:schroedingerQM} 
\end{equation}
would resemble the solution of the classical KvN wavefunction $\psi(x,t)$, provided that they are prepared with identical initial states. In other words, if $\psi(x,0)=\psi_\text{QM}(x,0)$, the solution of $\psi$ following Eq.~\eqref{eq:Schroedingeri} and the solution of $\psi_\text{QM}$ following Eq.~\eqref{eq:schroedingerQM} are identical, $\psi\l(x,t\r)=\psi_\text{QM}\l(x,t\r)$ $\forall t\ge 0$. As such, on the quantum computing side, the goal is to devise a quantum circuit which implements the finite-time propagator, $\exp\l(-i {H}_\text{QM} t\r)$  \cite{berry2007EfficientQuantumAlgorithms}.

Note that the nonlinearity now is hard-coded in the Hamiltonian operator $H_\text{QM}$. The evolutionary equation is still linear in $\psi_\text{QM}$. This property is inherited from the Liouville picture, that the operator $\mathcal{L}^\dagger$ encodes the nonlinear flow in the phase space.

The advantage of this approach is that the structure of the finite-time propagator $\exp\l(-i H_\text{QM} t\r)$ is unitary. This formulation is now fundamentally consistent with quantum mechanics. We now show two different ways to simulate the above process Eq.~\eqref{eq:schroedingerQM}, depending on the operational state space.

Suppose one can identify a quantum system whose state space coincides, or reasonably approximates, the continuum state space of the dynamical system $\mathbb{R}^N$. In such a system, suppose we can construct the momentum operator $\hat{P}$, the flow operator $F\l(\hat{X}\r)$, and therefore the Hamiltonian $H$. All the operators now are physical quantities of the quantum system, so there is a possibility that we can devise a physical system to ``mimic'' the Hamiltonian. Then, we can simply evolve the quantum system forward in time to solve the nonlinear dynamical system. We refer to this approach as a ``analog quantum simulation'' approach, in which, $\hat{P}$ is the true physical momentum in the continuum state space. One potential pitfall of this approach is that it is not possible to represent a continuous $\delta$-distribution on the discrete Hilbert space of an analog quantum simulator.
\black{A second potential pitfall is that not all allowable initial states of the discrete Hilbert space might be initialized on an analog devices due to limitations of control.} 
Finally, there is always an uncertainty associated with an initial condition that could lead to inaccurate prediction. For instance, if the phase-space flow diffuses the probability mass in some direction. 

A perhaps more realistic approach is to use gate-based QC, whose fundamental units are discrete quantum states.~\cite{somma2016quantum}. In this case, we would need to discretize the state space $\mathbb{R}^N$ into discretized points, and approximate the classical KvN wavefunction $\psi$ by a wavefunction with discrete support on a QC. Importantly, now the momentum operator $\hat{P}=i\partial_x$ can only be approximated with discrete-state operations. This induces numerical artifacts due to finite-size discretization; see the illustrative example below. 

One would also need to be able to implement an approximate flow operator $F\l(\hat{X}\r)$ to construct the Hamiltonian  \cite{park2017qubit,welch2014efficient}. Finally, we use one of the previously mentioned methods \cite{sornborger1999higher,berry2015simulating,low2019hamiltonian} to construct the finite-time propagator to evolve the system assuming the constructed Hamiltonian is sparse. 

\begin{figure}[!t]
\centering
    \includegraphics[width=1.0\columnwidth]{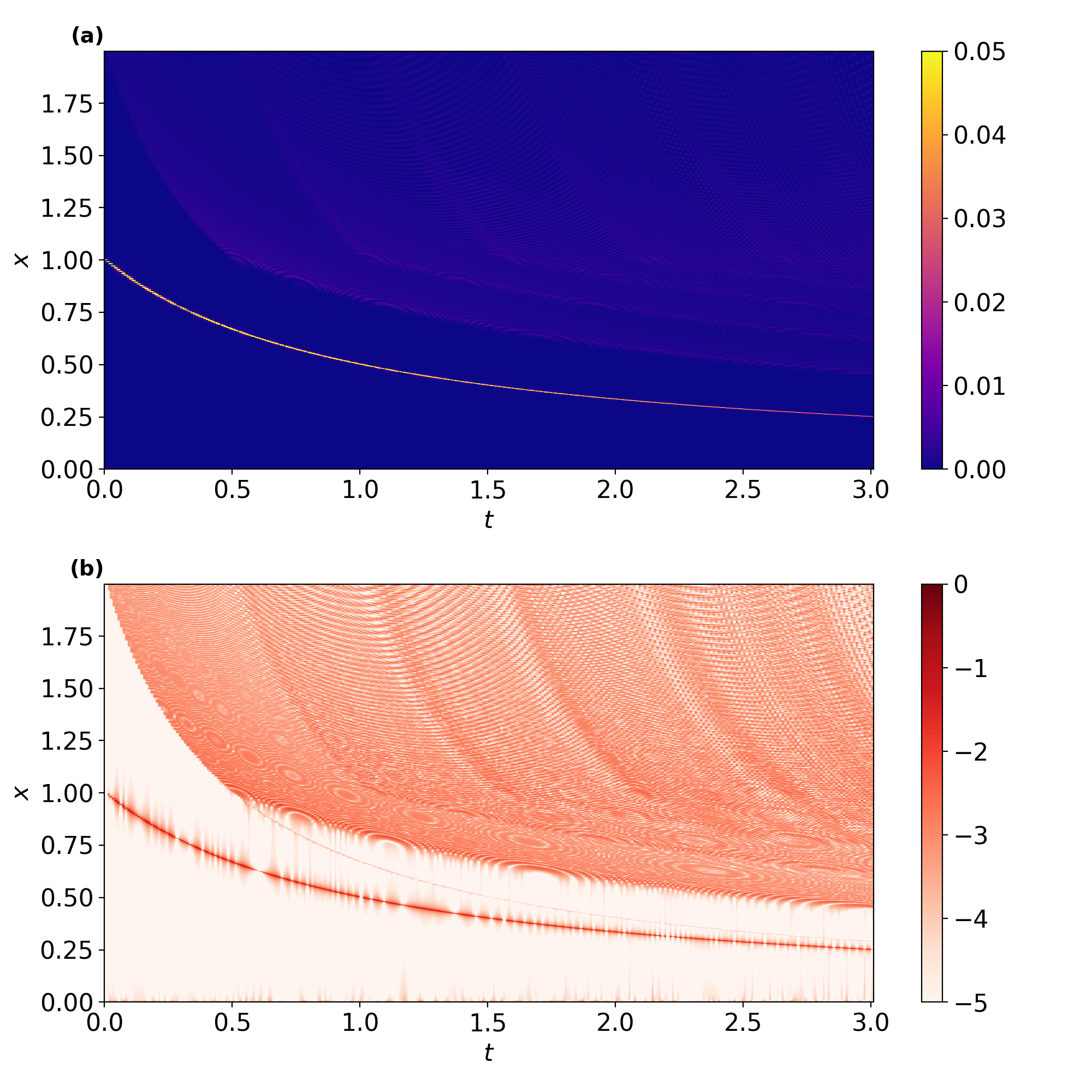}
    \caption{The probability density predicted by KvN mechanics with 10-qubit discretization ($2^{10}$ points in $x\in[0,2.0]$.) Panel (a) shows the evolution of the probability density $\l\vert \psi\l(x,t\r) \r\vert^2$, visualized as a heatmap. $\l\vert\psi\r\vert^2 \ge 0.05$ is considered as saturated to increase the contrast of the image. The mode of the distribution evolves almost identically with the deterministic solution (cf.~analytical solution in Figs.~\ref{fig:carleman} and \ref{fig:KvN-summaryStatistics}). (b) We plot  $\log_{10}\l\vert \psi\l(x,t\r) \r\vert^2 $ as the heatmap to reveal the fine ripple structure resembling a Gibbs-like phenomenon due to the finite dimensional derivative operator.}
    \label{fig:KvN}
\end{figure}

\begin{figure}[!t]
\centering
    \includegraphics[width=1.0\columnwidth]{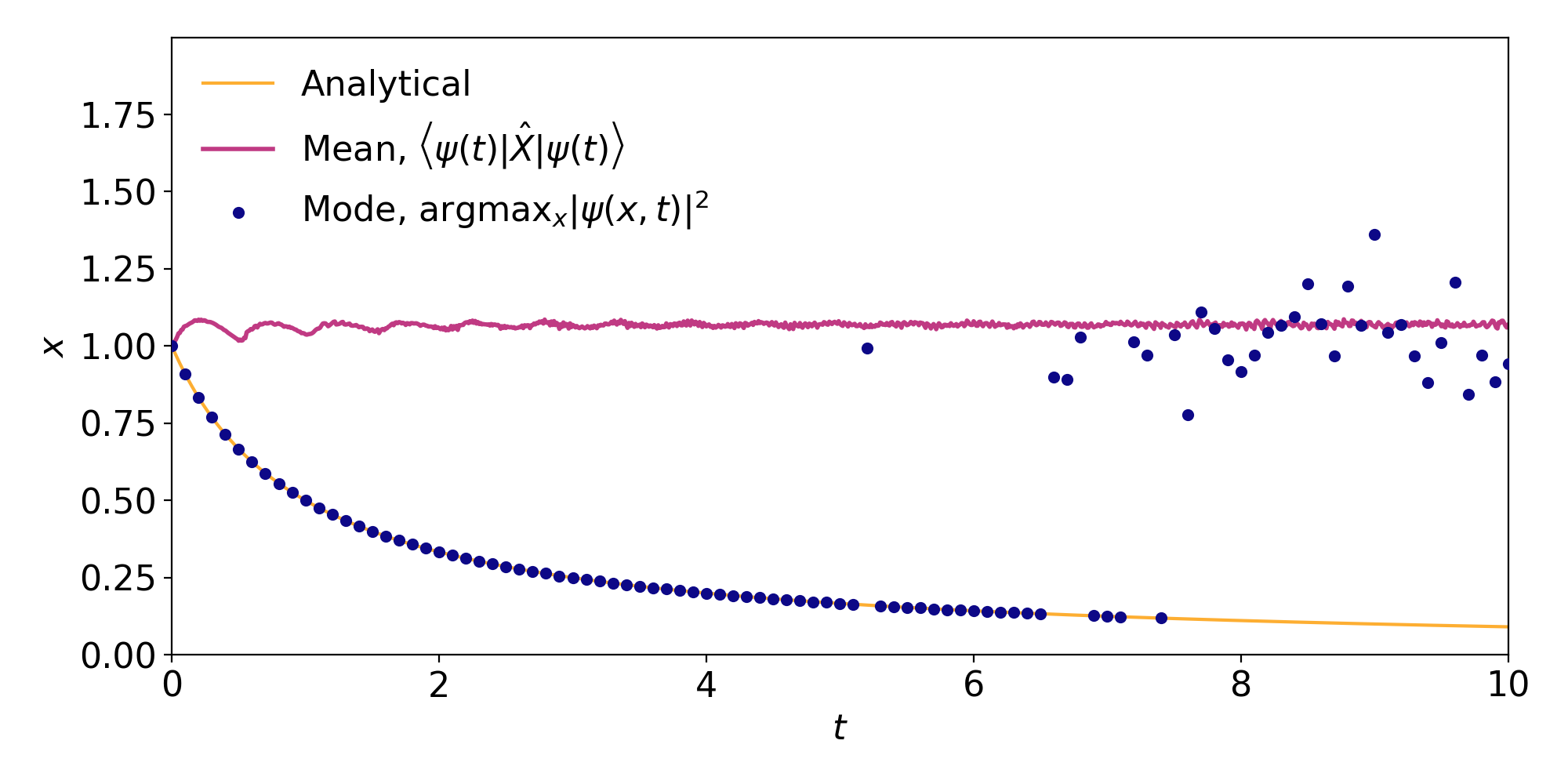}
    \caption{Summary statistics shows that the mode of the probability distribution follows the analytic solution over a short time horizon, $t\lesssim 5$, beyond which the Gibbs-like phenomenon is large enough to disturb the mode into almost random fluctuations. The mean, $\l\langle  \psi(t) \vert \hat{X} \vert \psi(t)  \r \rangle $, on the other hand, is not a useful predictor for all $t\ge 0$.}
    \label{fig:KvN-summaryStatistics}
\end{figure}

\emph{Example.} To illustrate potential problems and pitfalls, we again consider the stylized $\dot{x}=-x^2$. We consider a finite domain $x\in\l[0,2\r]$, and only a simple initial condition $x(0)=1$. We partition the system into $2^{10}$ equidistant spatial points and use a discretized wavefunction $\psi=[\psi_1, \psi_2, \ldots \psi_{1024}]^T.$ The system can be represented as a $10$-qubit quantum state. 

We coded the flow operator $\widehat{F(X)}=\hat{X}^2$ and $\hat{P}$ operator in the position representation with an FFT-based (Fast Fourier Transform) differentiation. Existing quantum Fourier transformation algorithms enable direct implementation of such FFT-based construction of the $\hat{P}$ operator. After constructing ${H}_\text{QM}$ on this discretized lattice, we use \texttt{scipy.linalg.expm} to compute a finite-time propagator $\hat{U}_\delta:=\exp(-i {H}_\text{QM} \delta)$, with $\delta=0.01$. Again, this is the optimal setting, that we aim to use the most accurate propagator to reveal the finite-size lattice effect. We then use $\hat{U}_{0.01}$ to propagate the wavefunction $\psi$ to $t=10$ iteratively (1000 time steps of size $\delta$). 

Figure \ref{fig:KvN}(a) shows the evolution of the probability density $\vert \psi \vert^2$ up to $t=3$. Although the majority mass of the distribution evolves according to the analytical solution of $\dot{x}=-x^2$, noticeable ripples form above the solution $x(t)$. To highlight these ripples, we plot the logarithms of the probability density in Fig.~\ref{fig:KvN}(b). Resembling the Gibbs' phenomenon, these ripples should not be surprising. The eigenfunctions of $\hat{H}$ are all oscillatory in time, but the true dynamics ($\dot{x}=-x$) is dissipative. As a finite discretization restricts us to only $2^{10}$ basis functions, it is not surprising that after a certain time, we lose the ability to use only $2^{10}$ oscillatory modes to approximate and predict long-term behavior. A more thorough analysis of the underlying cause of the Gibbs-like phenomenon can be found in 
Appendix \ref{app:1}.

In Fig.~\ref{fig:KvN-summaryStatistics}, we show that the mode, defined as the lattice point which has the largest amplitude of the wavefunction, evolves almost according to the analytic solution, but only within a finite horizon $t\lesssim 5$. Beyond this horizon, the effect of Gibbs-like phenomena is too strong and we would no longer be able to use the location of the largest amplitude to predict the dynamics. 

Interestingly, the expected value of the $\hat{X}$ operator behaves extremely poorly even when the effect of the Gibbs-like phenomenon is small. We remark that all observations are robust to algorithmic parameters, such as the choice of the location of the boundary, number of discretization points, etc., see Supplemental Figs.~\ref{fig:KvN-BC1} and \ref{fig:KvN-BC2} with different boundary settings. When changing the initial distribution from $\delta$-like (i.e., probability mass $=1$ on only a specific discretization point) to Gaussian-like, the Gibbs phenomenon is delayed and the prediction horizon by either the mode and expectation value can be improved; see Supplemental Fig.~\eqref{fig:KvN-Gaussian}.

\subsection{Directly solving Liouville equations}

\label{sec:solveLiouville}

The fact that KvN mechanics requires discretization which induces numerical artifacts motivates us to reconsider the problem---if one needs to discretize the space anyway, why not simply come back to the very fundamental problem of solving the Liouville equation by discretization? We propose to discretize the space with countably many points, and formulate the probabilities of the system at each point as a joint distribution with discrete support. A stochastic process can be formulated and described by Chemical Master Equations (CMEs). The joint probability distribution converges to the Liouville equation \eqref{eq:Liouville} in the continuum limit via the well-known Kramers--Moyal expansion. The advantage of the CMEs is their transparent resemblance with the original dynamical system. Using CME's do have a few drawbacks. First, similar to the KvN mechanics, for any finite number of discretization points, intrinsic noise would be introduced and one would need to simulate multiple paths for collecting the statistics. Secondly, the forward propagator $\mathcal{L}^\dagger_{\Delta}$ (a discrete approximation of the Liouville operator $\mathcal{L}^\dagger$ with spatial grid size $\Delta$) is not a unitary operator, and thus, one cannot simply use the elementary unitary operations in QC for integrating the system forward in time. Instead, we would need to resort to those QC algorithms for solving the general linear equations, noting that CMEs are linear in the variables $p_j(t)$. In these probabilistic algorithms, additional sampling is needed to determine the solution. Nevertheless, the numerical accuracy of this discretization scheme seems to be much better than the KvN mechanics. 

We finally remark that in terms of the Chemical Master Equations, the adjoint of the forward operator $\mathcal{L}_\Delta := \l(\mathcal{L}_{\Delta}^\dagger\r)^\dagger$ is the Koopman operator of the approximate process on the finitely many discretization points, and such an adjoint operator is an approximate Koopman operator. When the discretization is dense enough, the analysis of the approximate Koopman operator is similar to a approximate Koopman analysis using Radial Basis Functions \cite{williamsKernelbasedMethodDatadriven2015}, which have been empirically shown to be practical.

\begin{figure}[!t]
\centering
    \includegraphics[width=1.0\columnwidth]{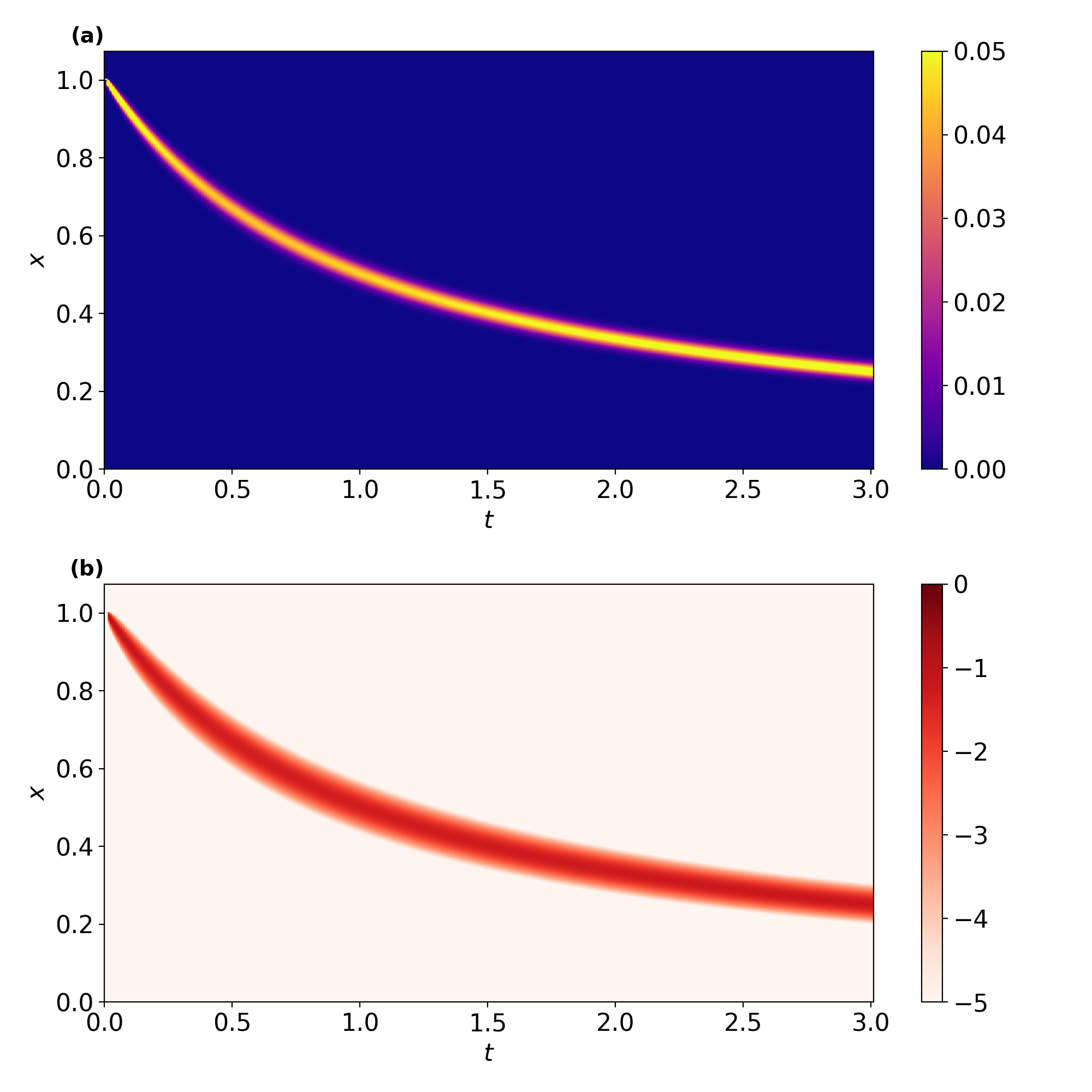}
    \caption{The probability density predicted by the Chemical Master Equation with $2^{10}$ discretization points on a domain $x\in\l[0,2\r]$. Panel (a) shows the evolution of the joint probability distribution $\mathbf{p}(t)$, visualized as a heatmap. $p_i(t)\ge 0.05$, $i=1\ldots 1024$ is considered as saturated to increase the contrast of the image. We plot $\log_{10} \mathbf{p}(t)$ as the heatmap in Panel (b), for making comparison to KvN mechanics Fig.~\eqref{fig:KvN}(b).}
    \label{fig:Liouville}
\end{figure}

\begin{figure}[!t]
\centering
    \includegraphics[width=1.0\columnwidth]{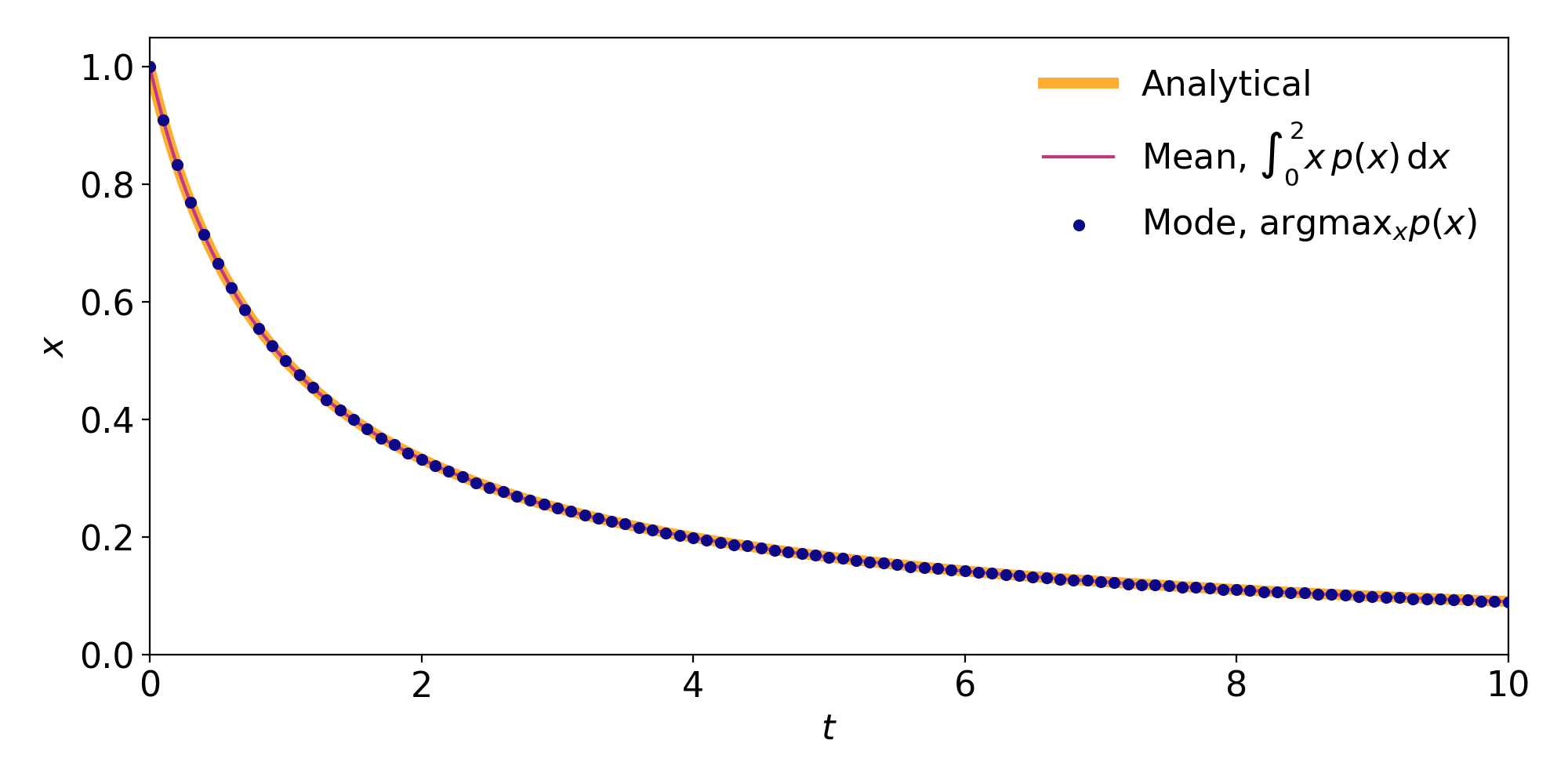}
    \caption{Summary statistics shows that both the mode of the probability distribution and the expectation value follows the analytic solution, showing that the Liouville discretization scheme performs better than the KvN mchanics Fig.~\ref{fig:KvN} on this problem.}
    \label{fig:Liouville-summaryStatistics}
\end{figure}

\noindent{\emph{Example.}} In terms of our example $\dot{x}=-x^2$, we discretize the space $x\in[0,2]$ into 1024 equidistant points $\l[x_1, \ldots, x_{1024}\r]^T$. We denote the grid size by $\Delta$. The probabilistic system now is captured by 1024 probabilities, $\mathbf{p}(t):=\l[p_1(t), \ldots, p_{1024}(t)\r]^T$.
Since $p(t)$ obeys the linear Liouville equation, we can solve it using the quantum algorithms for linear differential equations~\cite{berry2014HighorderQuantumAlgorithm,berry2017quantum}. These algorithms would then output a quantum state whose amplitudes are proportional to the probabilities, i.e. $|\psi\rangle\propto \sum_n p_n(t) |n\rangle$.~\footnote{Note that here the amplitudes of the quantum state are proportional to the classical probabilities evolving according the Liouville equation, whereas in KvN the absolute value squared amplitudes, i.e. quantum probabilities, were proportional to the classical probabilities. As a result, here, the information about the nonlinear system cannot be expressed as an expectation value. Nevertheless, there are ways to extract such information.} Then, the forward operator $\mathcal{L}_\Delta^\dagger$ is a sparse Markov matrix, whose entries are defined as
\begin{equation}
\label{eq:CMEL}
    \l[\mathcal{L}_\Delta^\dagger\r]_{i,j}= \l\{\begin{array}{cl}
     -x_j^2 \Delta^{-1} & \text{if } j>0 \text{ and } i=j, \\
     x_j^2 \Delta^{-1} & \text{if } j>0 \text{ and } i=j-1, \\
     0 & \text{else.}
    \end{array}\r.
\end{equation}
We can think of the state variable $\mathbf{p}(t)$ as the joint probability distribution of the position of a continuous-time random walk on the lattice. At any time $t$, the walker at $x_j$ is only allowed to move to the discretization point to its left $x_{j-1}$ with a transition rate $-x_j^2$. Then, the CME can be written as a set of 1024 ordinary differential equations, $\mathbf{p}(t)=\mathcal{L}_\Delta^\dagger \cdot \mathbf{p} (t)$. Again, we used \texttt{scipy.linalg.expm} to calculate the propagator $\exp\l(\delta \mathcal{L}_\Delta^\dagger\r)$, with $\delta = 0.01$. The initial condition is set as $\delta$-distribution-like, i.e., $p_i(t)=1$ for $x_i=1$ and $0$ otherwise. Analogous to Figs.~\ref{fig:KvN} and \ref{fig:KvN-summaryStatistics}, we visualize the probability densities in the linear and $\log_{10}$ scale as heat maps in Figs.~\ref{fig:Liouville} and \ref{fig:Liouville-summaryStatistics}, showing the improved accuracy for the problem. 

We remark that the improvement of the CME approach is specific to the $\dot{x}=-x^2$ model. In a parallel analysis on the Van der Pol oscillator presented in Appendix \ref{app:2}, we observed that with a finite discretization in phase space, the CME inevitably introduces phase diffusion to the oscillatory dynamics. The KvN mechanics, whose fundamental building blocks are also oscillatory, was seen to perform marginally better than the CME on the Van der Pol oscillator.

The random walk and the corresponding CME may seem to be an \emph{ad hoc} and heuristic construct for solving the Liouville equation, with the guarantee that the solution of the CME asymptotically (on the infinitely dense grid) converges to the solution of the Liouville equation through the Kramers--Moyal expansion. Another point of view is that the Liouville equation \eqref{eq:Liouville} is a hyperbolic conservation law and the CME (\ref{eq:CMEL}) corresponds to its first-order upwind finite-volume discretization (see, for example, \cite{leveque_finite_2002}). Specifically, we can think of an explicit Euler scheme for solving the Liouville equation as a conservation law for  $\rho^n_i:=\rho(x_i, n\delta) $, evolved over the time step $\delta \ll 1$ as
\begin{equation}
\label{eq:FOUphi2}
    \rho_i^{n+1} = \rho_i^n + \delta \Delta^{-1}
      \left(\phi_{i+1/2}^n - \phi_{i-1/2}^n\right),
\end{equation}
where $\phi_{i+1/2}^n$ is the flux passing through the interface located at $x_{i+1/2}:=\l(x_{i}+x_{i+1}\r)/2$ at time $t=n\delta$. We can choose a specific first-order upwinding scheme $\phi_{i+1/2}^n:= x_{i+1}^2 \rho_{i+1}^n$
or using (\ref{eq:CMEL}), we can write this method as a CME with forward-Euler time integration:
\begin{equation}
    \rho_i^{n+1} = \rho_i^n + \delta
      \left\{\l[\mathcal{L}_\Delta^\dagger\r]_{i,i+1} \rho^n_{i+1} +  \l[\mathcal{L}_\Delta^\dagger\r]_{i,i}\rho^n_{i}\right\}.
\end{equation}
This discretization is well known to be monotone and thus will not exhibit Gibbs-like numerical artifacts.  However, higher-order accurate extensions will result in Gibbs, unless nonlinear feedback is implemented (see \cite{leveque_finite_2002} for more details), and introducing nonlinearity into the numerical method works against our goal of implementation on a QC.

\section{Discussion and Summary} \label{sec:discussion}

We have presented two major motivating results that have driven this article. First, we have provided a unified theoretical framework, a ``roadmap'', tying together various propositions that can be used to simulate nonlinear dynamical systems on a quantum computer. We believe that classifying and pinpointing potential methods on the ``roadmap'' is extremely valuable not only to the field of quantum computation, but also to the field of classical dynamical systems. On the one hand, quantum computation could leverage recent advancements in the classical domain, such as various variants of the Dynamic Mode Decomposition and the revived Koopmanism. On the other, classical computational dynamics could leverage the high-dimensional space and faster algorithms that are only feasible in the realm of quantum computing. Our second motivation is to point out potential pitfalls and major technical bottlenecks of these methods in order to provide some guidance for identifying more and less viable directions for further research on quantum algorithms for simulating classical systems. 

Both Koopman von Neumann wave mechanics \cite{bogdanov2014StudyClassicalDynamical,bogdanov2019QuantumApproachDynamical,joseph2020KoopmanNeumannApproach} and Carleman linearization \cite{carleman1932,kowalski1991NonlinearDynamicalSystems,liu2021efficient} have been proposed for the purpose of simulating classical nonlinear dynamics on a quantum computer, leveraging their linear representations.
Towards the first aim, we present the duality between the forward representation and backward representation of dynamical systems. 
We identify a connection between the forward Perron--Frobenius representation and the Koopman von Neumann wave mechanics theory. We also make a connection between the backward Koopman representation and Carleman linearization.  We also discuss why the direct nonlinear integration of the flow is not feasible on a quantum computer. 
Although these are not new theories, we attempted to make the duality clear by juxtaposing the representations in a single article, as most published papers are anchored on only one representation.

With the goal of using Koopman von Neumann wave mechanics to simulate nonlinear systems, we identified that it is a forward, Perron--Frobenius representation, despite its mysterious references to the papers by Koopman and von Neumann, which make use of the backward representation. We discovered that this method is similar to the Proper Orthogonal Decomposition in the classical domain, that the fundamental modes (basis functions) of the dynamics are mutually orthogonal in space, and the eigenvalues are all purely imaginary, leading to wave-like dynamics. We pointed out that the major technical difficulty lies in the fact that discretization can lead to a Gibbs-like phenomenon caused by the under-resolved dynamics on length scales smaller than that of the discretization. Numerical methods that suppress Gibbs are typically nonlinear, even if the original equation being discretized is linear, which is counter to our original goal of solving a linear system on a QC.

Unfortunately, many classical systems, including Hamiltonian ones with chaos, have the feature that the dynamics excites small-scale features in the phase space. Perhaps the most famous example is the Kolmogorov cascade in turbulence. Simulating such systems may require arbitrarily fine discretizations and the question of whether quantum computing can lead to efficiency gains over classical nonlinear integrator operating in the phase-space remains open \cite{lewis2023limitations}.

As for the proposition of leveraging Carleman linearization to simulate nonlinear dynamics, we identified that Carleman linearization is a special case of the Koopman representation, in which one uses polynomial functions as the basis functions spanning the Hilbert space. While Liu et al.~\cite{liu2021efficient} showed that it is possible to use Carleman linearization, it is important to point out the limitation that only a special class of dynamical systems is considered in the study: the nonlinearity of the flow example is quadratic, and more importantly, the class of models imposed very strict conditions on the linear stability, i.e. that the system is stable in every direction in the phase space (all eigenvalues of the linearized dynamics are strictly less than zero). As such, any systems with saddles, unstable fixed points, stable fixed points that are the result of nonlinearity, or limit cycles would not fall into this class. For example, both our toy examples, $\dot{x}=-x^2$ and the Van der Pol oscillator $\ddot{x} - 0.5 \l(1-x^2\r) \dot{x} + x =0$, are not in the class Liu et al.~ analyzed \cite{liu2021efficient}. 

The major problem of using Carleman linearization manifests in our numerical illustrations and Appendix \ref{app:carleman}, that there is only a short time horizon during which the linearization would lead to the correct solution. The strict requirement of linear stability in Liu et al.~\cite{liu2021efficient} ensures that errors  induced by truncation or projection decay away. For general nonlinear systems, we identify that the major difficulty lies in the identification of closure or projection schemes which are linear in the finite set of functional basis. Unfortunately, such schemes are highly system-dependent and generally unknown \emph{a priori} without a thorough understanding of the system. Through numerical simulations, we showed that a scheme does exist for the simple equation that we study and can be learned from simulation data via the Extended Dynamic Mode Decomposition \cite{williamsDataDrivenApproximation2015}. 

Recently, Amini et al.~\cite{amini2021ErrorBoundsCarleman} showed that the error of the Carleman linearization can be bounded exponentially by the truncation order $N$, but this statement does not hold for any arbitrary time. As is clearly described in the paper, provided with the assumption that the flow field can be approximated by a Maclaurin expansion, the error for a sufficiently short horizon is exponentially bounded. However, an estimate of the horizon was not provided, but is discussed in \cite{forets_explicit_2017}. As is shown in our simple numerical examples, undesirable errors can be detrimental in the long time limit even for the stable $\dot{x}=-x^2$ system. Conditions for finite-time blowup of Carleman are discussed further in Appendix \ref{app:carleman}. Thus, it is unlikely that adoption of Carleman linearization for simulating general dynamical systems will be useful when the target is not short-horizon prediction. Furthermore, the number of polynomial basis functions has a stiff scaling with the dimensionality of the phase space and the order of the polynomial (sum of the exponents).

In modern data-driven Koopman learning for dynamical systems, we have learned the valuable lesson that it is challenging to control high-order polynomial basis functions (imagine $x^{100}$ for some $x>1$!) There have been propositions for using radial basis functions or Hermite polynomials \cite{williamsDataDrivenApproximation2015,chorin2002optimal} as the basis functions. 

Importantly, it is also well-perceived that the choice of basis functions and the quality of the finite-dimensional approximation strongly depends on the statistical properties (e.g., the ergodic measure) of the specific dynamical system, because the inner product in the Hilbert space $L^2 \l(\mathbb{R}^N, \dd \mu \r) $ is tied to the statistical properties of the dynamics. Such knowledge can be potentially leveraged to design algorithms for simulating specific dynamical systems on a quantum computer. 

If one pursues this route for Koopman representations, for those systems which have finite-dimensional Koopman invariant spaces, knowing the finite set of basis functions which span the invariant subspaces leads to a (potentially nonlinear) transformation of the nonlinear dynamics to a linear and closed one in the Hilbert space, fully solving the problem. 

Perhaps the most famous example to this is the Cole--Hopf transform of the nonlinear Burgers' equation. To this end, we identify the technical difficulty of using a quantum computer lies in the implementation of general nonlinear transformations between the phase space and the observable space \cite{holmes2023nonlinear}. For those systems without finite-dimensional subspaces, how one could improve finite-dimensional approximations for  infinite-dimensional dynamics of observables by selecting the optimal set of basis functions remains an open problem. 

With our numerical illustrations, we aim to show the simple fact that despite the nice linear properties in these representations---the forward Liouville/Koopman von Neumann mechanics or the backward Koopman representation, there is a trade-off that the operational Hilbert spaces are infinite dimensional. Any reduction to finite dimension inevitably induces numerical errors and the prediction of the dynamics can only be accurate before a finite horizon. \emph{There is no free lunch: changing the representation merely changes the formulation of the problem without solving these issues.} 

In Koopman von Neumann mechanics, a discretization into a finite number of grids in the phase space leads to a finite number of oscillatory modes which are not capable of resolving dynamics below a critical length scale, leading to a Gibbs-like phenomenon. 

Lloyd {\it et al.} \cite{lloyd2020quantum} takes a different approach to solving nonlinear differential equations based on the simulation of nonlinear Schr\"odinger equations. This entails designing an interacting many-body system evolving according to the standard Schr\"odinger equation such that the dynamics of each of the identical individual particles is described by a nonlinear Schr\"odinger equation in the thermodynamic limit (infinitely many particles). Finite many-body systems of this sort can be efficiently simulated with quantum computers. However, note that for the nonlinear Hartree equation, Gokler \cite{gokler_mean_2020} derived an upper bound on the error of the mean field approach that for a finite number of particles grows exponentially in time.  This behavior of the upper bound is analogous to that for the truncated Carleman approach \cite{forets_explicit_2017}.  Whether in general the mean-field approach has a corresponding finite-time horizon for accurate solutions is a subject for future work. All of the methods considered here, however, have similar issues related to discretization of some sort. Since the method of \cite{lloyd2020quantum} does not evolve observables, it is evidently a Perron-Frobenius method. Moreover, it has a significant advantage over the Perron-Frobenius methods discussed in this work (KvN and the direct method of Sections \ref{sec:KvN} and \ref{sec:solveLiouville}) because the encoding of the mean-field is much more efficient and could allow for exponentially finer lattices for fluid simulations.

In the Carleman and the more general Koopman representation, the choice of a finite number of basis functions leads to inevitable projection or truncation of the exponentially resolved observable dynamics. Without a data-driven approach to learning the projection operator from data, na\"ive truncations can even lead to diverging predictions.

Notwithstanding the above considerations, our perspective on using quantum computing to simulate nonlinear dynamical systems remains conservatively optimistic for two reasons. First, we are hopeful that the infinite-dimensional spaces can be reasonably approximated by a quantum computational system, whose state space is also a Hilbert space whose dimension is exponential in the number of qubits. Secondly, we are hopeful that the more efficient quantum linear solver can significantly enhance the integration process in comparison to classical linear solvers. 

Challenges exist of finding and implementing general nonlinear transformations \cite{holmes2023nonlinear} for those systems with closed Koopman invariant spaces, or devising finite-state approximations and estimating their errors for general systems. The solutions to these challenges are likely to depend on which nonlinear system one is studying.

\section{Acknowledgements}
YTL appreciates useful discussions with Francesco Caravelli. Thanks also to David Jennings and Matteo Lostaglio for feedback on an earlier version of this document. This work was supported by the Beyond Moore's Law project of the Advanced Simulation and Computing Program at Los Alamos National Laboratory managed by Triad National Security, LLC, for the National Nuclear Security Administration of the U.S. Department of Energy under contract 89233218CNA000001. 

\bibliography{refs}

\appendix

\begin{figure}[!t]
\centering
    \includegraphics[width=1.0\columnwidth]{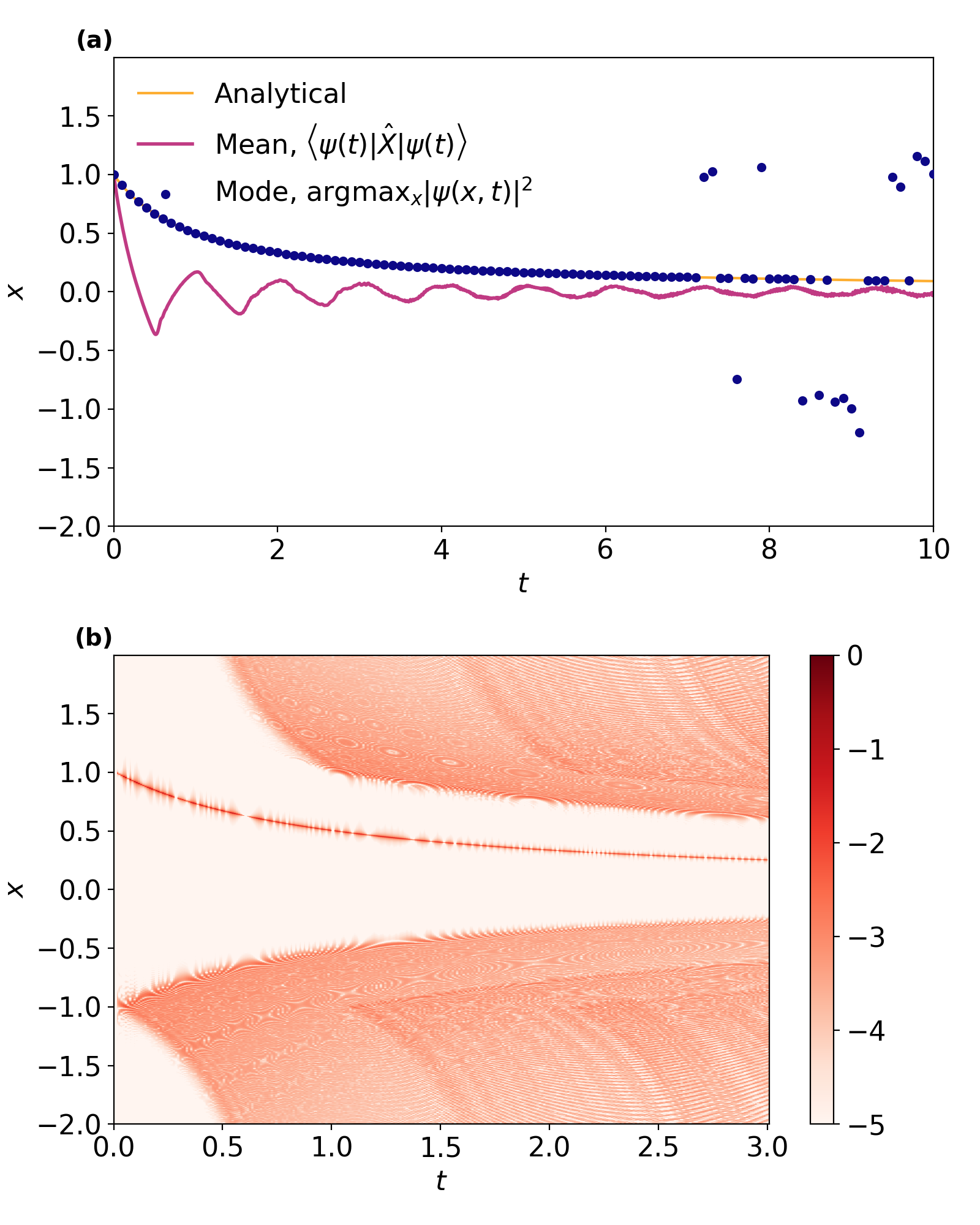}
    \caption{KvN mechanics for the $\dot{x}=-x^2$ model. The domain is extended to $[-2,2]$, with with $2^{11}$ discretization points. (a) the summary statistics (cf.~Fig.~\ref{fig:KvN-summaryStatistics}); (b) The base-10 logarithm of the probability density (cf.~Fig.~\ref{fig:KvN}(b)).}
    \label{fig:KvN-BC1}
\end{figure}

\begin{figure}[!t]
\centering
    \includegraphics[width=1.0\columnwidth]{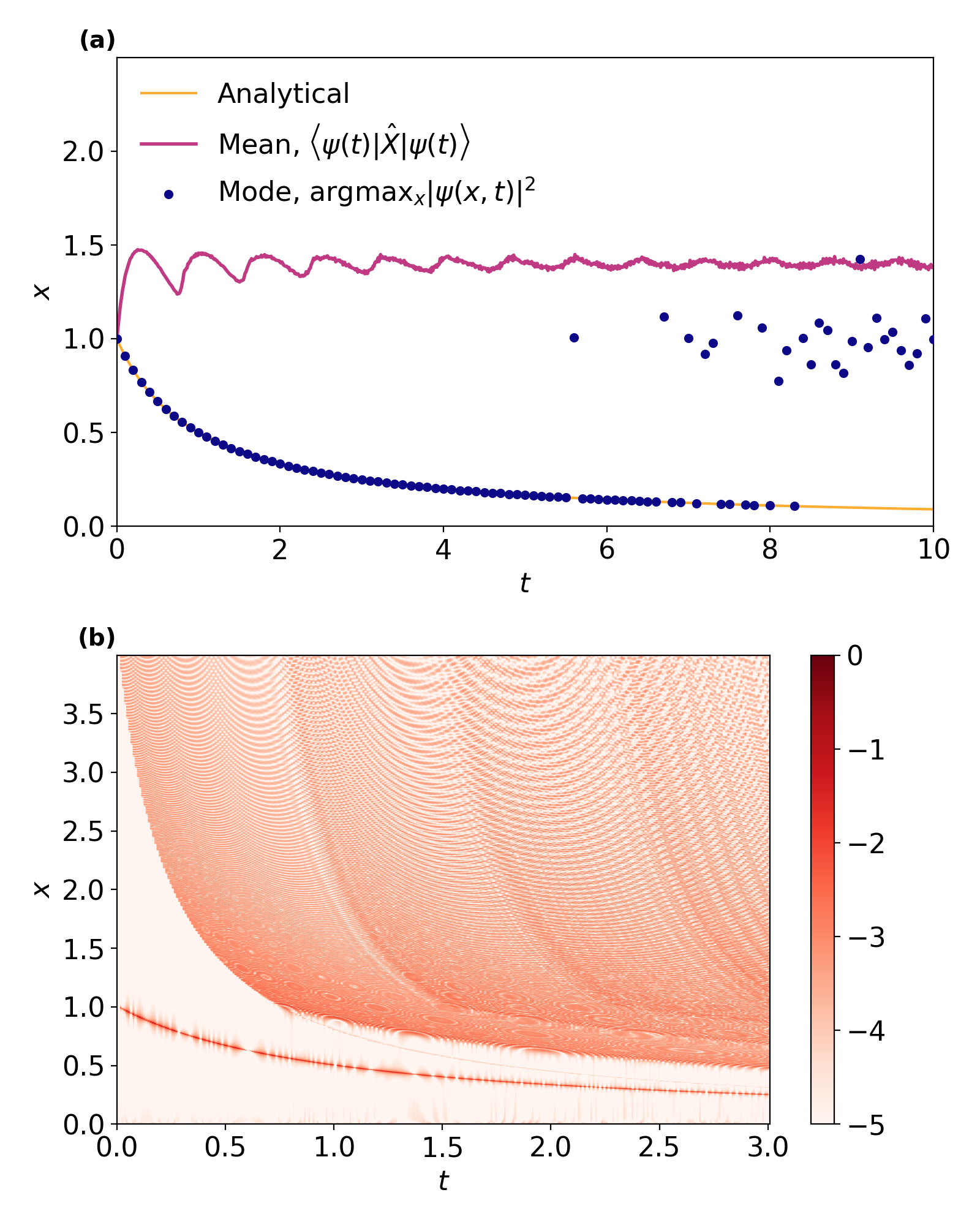}
    \caption{KvN mechanics for the $\dot{x}=-x^2$ model. The domain is extended to $[0,4]$, with $2^{11}$ discretization points. (a) the summary statistics (cf.~Fig.~\ref{fig:KvN-summaryStatistics}); (b) The base-10 logarithm of the probability density (cf.~Fig.~\ref{fig:KvN}(b)).}
    \label{fig:KvN-BC2}
\end{figure}

\section{Effects of boundary and initial condition in KvN mechanics} \label{app:1}

We wondered if the ripples (Gibbs-like phenomenon) could be resolved by changing the boundary conditions and/or initial distributions. Analyses were performed by changing the domain of the system from $x\in\l[0,2\r]$, specified in the main text, to $[-2, 2]$ and $[0,4]$. To accommodate the extended domains, in both cases, we discretized the system into $2^{11}$ points. The summary statistics (cf.~Fig.~\ref{fig:KvN-summaryStatistics}) and the probability density in the base-10 logarithmic space (cf.~Fig.~\ref{fig:KvN}(b)) are presented in Figs.~\ref{fig:KvN-BC1} and Figs.~\ref{fig:KvN-BC2}. We conclude that the observations presented in the main text are robust to the change of the boundary setting. 

\begin{figure}[!t]
\centering
    \includegraphics[width=1.0\columnwidth]{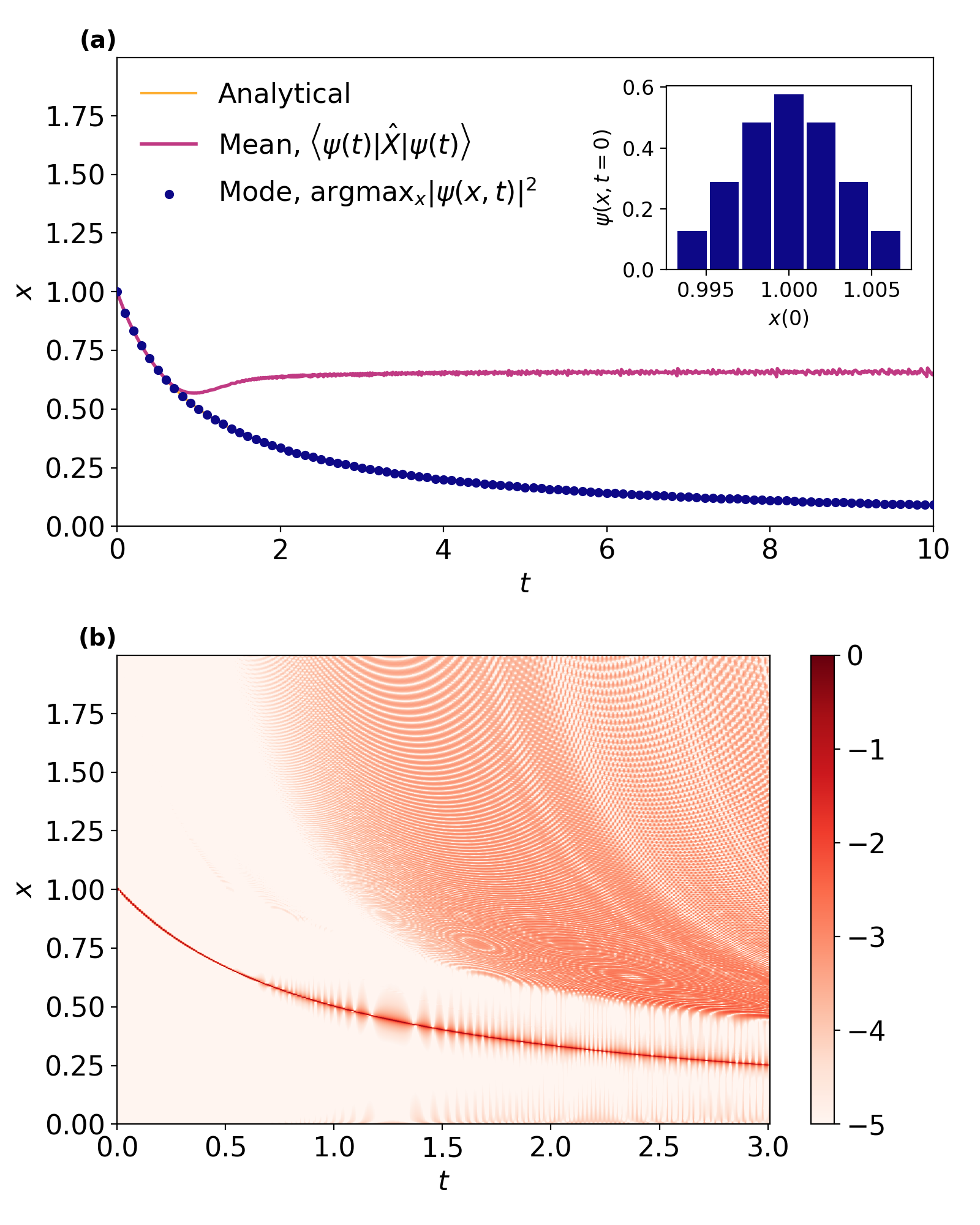}
    \caption{KvN mechanics for the $\dot{x}=-x^2$ model. Instead of $\delta$-like initial distribution, we put a Gaussian-like, 7-point distribution centered at $x=1$. (a) The summary statistics (cf.~Fig.~\ref{fig:KvN-summaryStatistics}). We plot the initial wavefunction in the inset; (b) The base-10 logarithm of the probability density (cf.~Fig.~\ref{fig:KvN}(b)).}
    \label{fig:KvN-Gaussian}
\end{figure}

\begin{figure}[!t]
\centering
    \includegraphics[width=1.0\columnwidth]{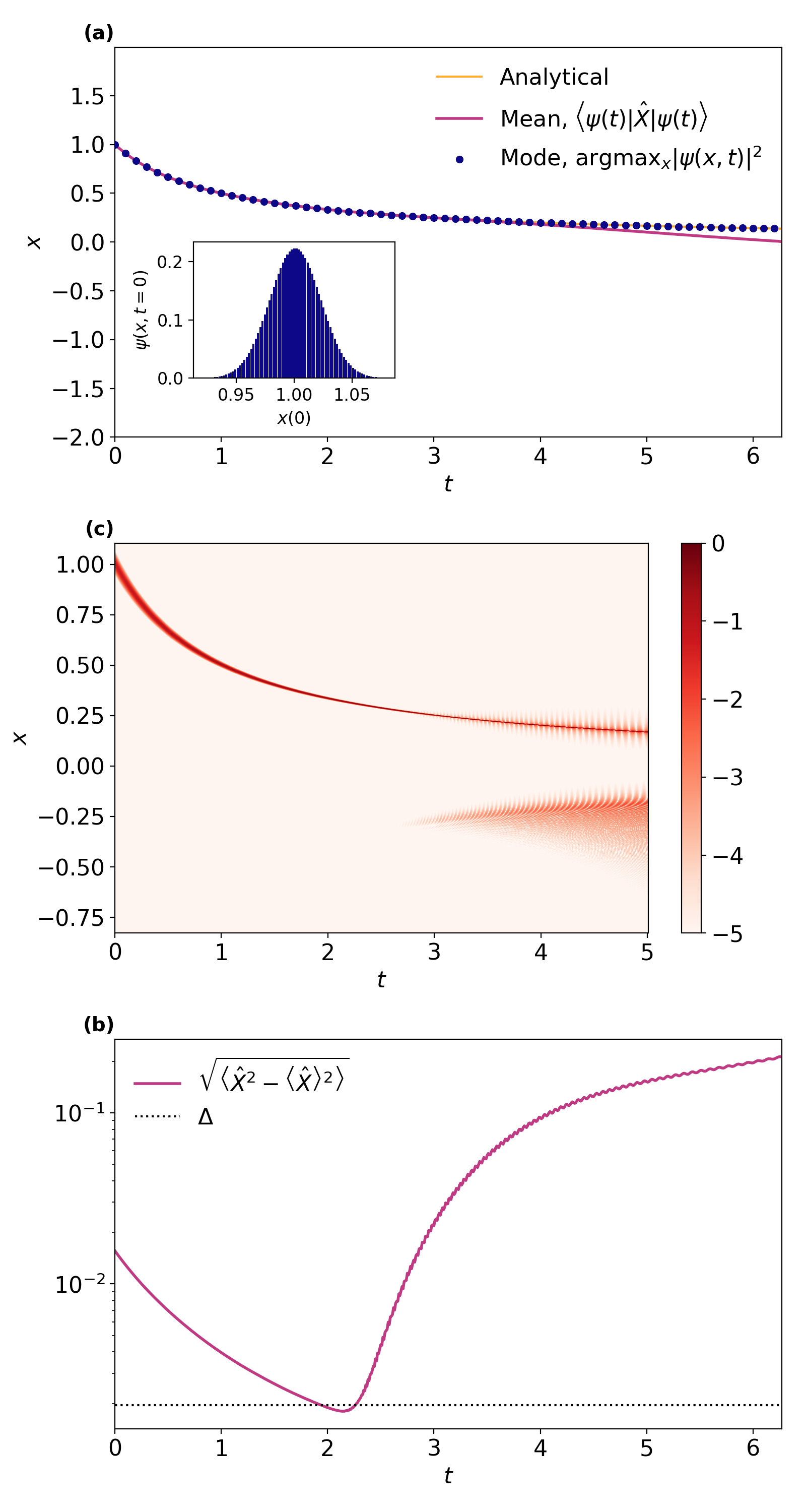}
    \caption{KvN mechanics for the $\dot{x}=-x^2$ model. Instead of $\delta$-like initial distribution, we put a Gaussian-like, 80-point distribution centered at $x=1$. (a) The summary statistics (cf.~Fig.~\ref{fig:KvN-summaryStatistics}). We plot the initial wavefunction in the inset; (b) The standard deviation of the distribution as a function of time; (c) The base-10 logarithm of the probability density (cf.~Fig.~\ref{fig:KvN}(c)).}
    \label{fig:KvN-Gaussian-long}
\end{figure}

To evaluate the effect of initial conditions, motivated by the physical constraint that we cannot initiate a $\delta$-distribution on a quantum system with continuum state space, we initiated a smoothed initial condition centered at $x=1$, shown in the inset of Fig.~\ref{fig:KvN-Gaussian}. We observed that the Gibbs-like phenomenon was delayed, leading to a short horizon ($t\le 1$) in which both the first moment and the mode of the probability distribution can reasonably approximate the analytical solution. The horizon of the mode prediction is extended to $t>10$ (Cf.~$t\approx 5$ in Fig.~\ref{fig:KvN-summaryStatistics}). With a wider initial distribution (inset in Fig.~\ref{fig:KvN-Gaussian-long}), we can achieve an even longer horizon before which the first moment and the mode of the discretized KvN wavefunction reasonably approximates the analytical solution, see the summary statistics in Fig.~\ref{fig:KvN-Gaussian-long}(a) and the joint probability distribution in the base-10 logarithmic scale in Fig.~\ref{fig:KvN-Gaussian-long}(b). In Fig.~\ref{fig:KvN-Gaussian-long}(c), we calculated the standard deviation of the distribution, showing the underlying cause of the Gibbs-like phenomenon. Because the dynamics $\dot{x}=-x^2$ has a global fixed point $x=0$, the probability distribution will eventually converge to this zero-measure point as $t\rightarrow \infty$. Thus, in the long-time limit, the majority of the probability mass is inevitably ``pinched'' into an interval that is smaller than the finite grid can resolve. We observe that soon after the standard deviation of the probability distribution decreases down to less than the grid-size $\Delta$, the solution became oscillatory: the standard deviation rebounded and started to increase. Noticeable Gibbs-like phenomena occurred soon after the rebound (see Fig.~\ref{fig:KvN-Gaussian-long}(b), $t\approx 3.5$) and destroys the predictions. 

We remark that the Gibbs-like phenomenon, well-known in spectral analysis \cite{gottlieb_numerical_1977,harten_multiresolution_1996,boyd_chebyshev_2000} is purely classical and due to under-resolved finite-dimensional projection of an infinite-dimensional space. 

\begin{figure}[!t]
\centering
    \includegraphics[width=1.0\columnwidth]{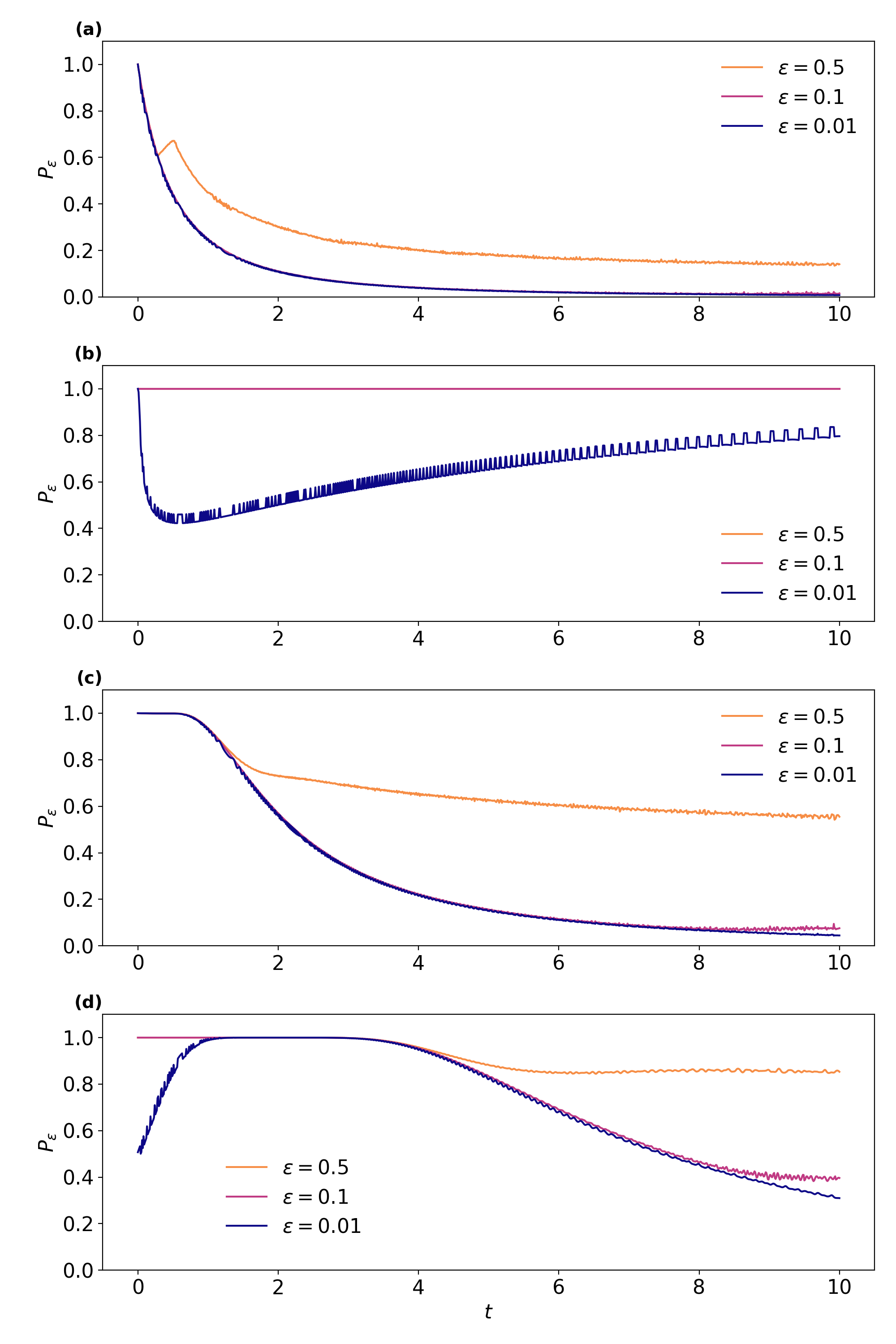}
    \caption{Error analysis: (a-d) correspond to Figs.~\ref{fig:KvN}, \ref{fig:Liouville}, \ref{fig:KvN-Gaussian}, and \ref{fig:KvN-Gaussian-long}. The integrated probability density in the interval $(x_\text{analytical}(t) - \varepsilon,x_\text{analytical}(t) + \varepsilon)$ is plotted against evolution time. Hence $p_\varepsilon$ can be interpreted as the probability of obtaining an answer within error $\varepsilon$ if a measurement in the computational basis is performed on the quantum state that encodes the solution to the dynamical system at that time.
    }
    \label{fig:error}
\end{figure}

We now observe an interesting trade-off: in comparison to a sharper (e.g., $\delta$-like) distribution, a wider initial distribution has a further prediction horizon but with a larger variance. To determine the summary statistics, e.g., the first moment of the probability distribution, the system with a wider initial distribution must require more samples. The observation seems to suggest that, at least for this system, one would choose the variance of the initial distribution to be such that at the prediction time, the variance of the evolving distribution is about $\Delta^2$. A larger than $\Delta^2$ variance at the prediction time would require additional samples to determine the summary statistics. A smaller than $\Delta^2$ variance would invoke the under-resolving problem. A certain optimization of the initial distribution, conditioned on the prediction time, is needed. 

Finally, we remark that the problem is generic if the dynamics has any ``pinching'', i.e., compression of the probability distribution in any direction in the phase space. Notice that even Hamiltonian systems may have such a feature: they are volume-preserving, but they can have a flow compressing in one direction and expanding in another direction. For those systems with invariant manifolds, such as fixed points and limit cycles, KvN mechanics with discretization does not seem to be a stable and scalable method for predicting long-time behaviors near those manifolds. 

\black{In the previous analysis we have focused on the mean and variance of the position observable in the KvN mechanics. Next, we will consider a different criterion for the quality of the simulation: the probability of outputting a position that is within $\varepsilon$ of the true value. This is given by the integrated probability density in the interval $(x_\text{analytical}(t) - \varepsilon,x_\text{analytical}(t) + \varepsilon)$, which is plotted against evolution time in Fig.~\ref{fig:error}. In practice, a measurement in the computational basis is performed on the quantum state that encodes the solution to the dynamical system at that time. The result in Fig.~\ref{fig:error} supports our previous conclusion that a larger variance in the initial distribution can improve the probability of measuring the position at the final time within a desired error.
}

\section{An alternative model: Van der Pol oscillator} \label{app:2}

To showcase that our observations are not conditioned on the particular choice of the model presented in the main text ($\dot{x}=-x^2$; see \S\ref{sec:solveLiouville}), we performed a parallel analysis on the Van der Pol oscillator. The evolutionary equations of the two-dimensional system are:
\begin{subequations}
\begin{align}
    \dot{x} ={}&  F_x(x,y):=y, \\
    \dot{y} ={}& F_y(x,y):= -x + \mu \l(1-x^2\r) y. 
\end{align}\label{eq:ODEvanDerPol}
\end{subequations}
The Van der Pol oscillator has a globally stable limit cycle, provided $\mu>0$. In this analysis, we fixed the model parameter $\mu=0.5$. The initial condition of the system is an arbitrary point near the stable limit cycle, obtained by integrating the system forward in time for a sufficient amount of time ($t=100$). We remark that the choice of $\mu$ led to a less stiff dynamical system. When we changed to $\mu=10$, which corresponds to a very stiff dynamical system, the methods below did not seem to solve the problem due to numerical instability. 

\noindent{\bf{Numerical solution}}. The Van der Pol oscillator does not have an analytical solution. To set up the baseline, we used the general-purposed \texttt{scipy.optimize.solve\_ivp} with \texttt{lsoda} to obtained the numerical solution. The absolute and relative errors are both set at a stringent $10^{-10}$. 

\begin{figure*}[!t]
\centering
    \includegraphics[width=0.75\textwidth]{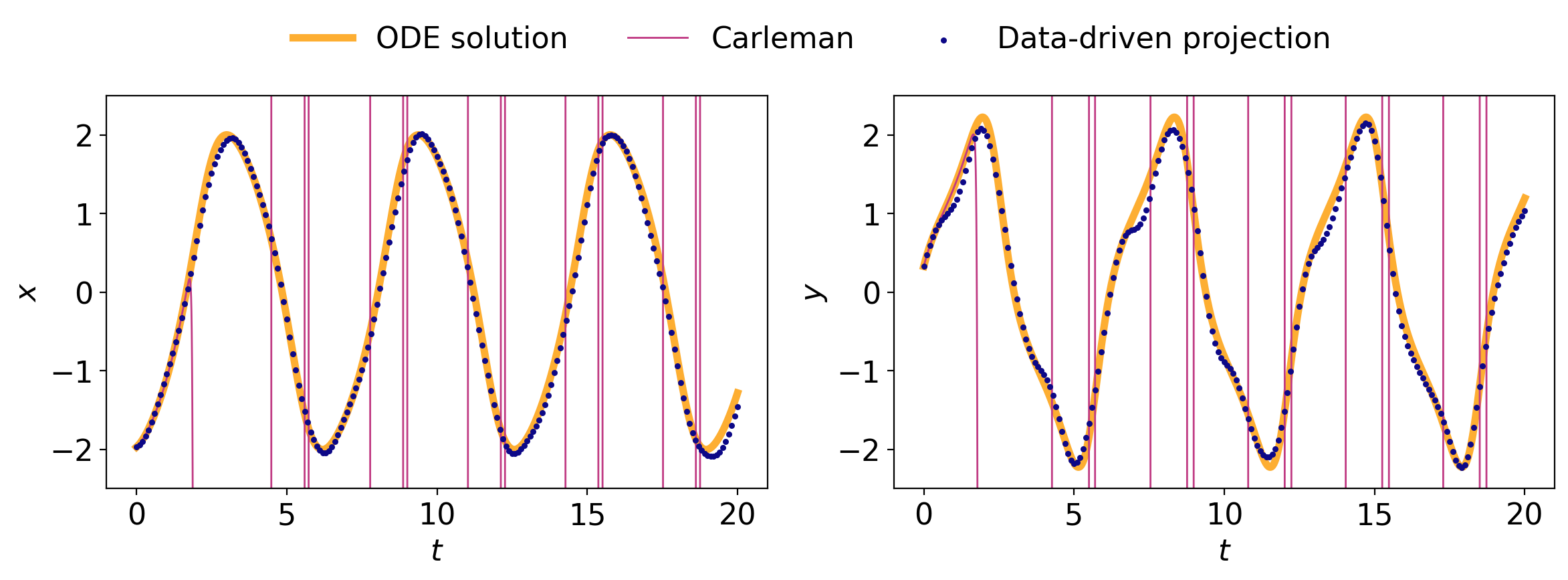}
    \caption{Predictions by the Carleman linearization and approximate Koopman operator learned from EDMD for the Van der Pol oscillator. }
    \label{fig:vanDerPol-Carleman}
\end{figure*}

\noindent{\bf{Carleman linearization.}} We perform the Carleman linearization on the set of polynomial functions $g_{m,n}:=x^m y^n$. It is straightforward to show that the evolutionary equations are:
\begin{align}
    \dot{g}_{m,n} ={}& m g_{m-1,n+1} - n g_{m+1,n-1} \nonumber \\
    {}&+ \mu n \l(g_{m,n} - g_{m+2, n}\r).
\end{align}
The truncation scheme is devised as follows. First, we only consider a polynomial order of $m+n\le49$. Thus, the state space leads to 1,275 functions. By observing the evolutionary equation \eqref{eq:ODEvanDerPol}, we know that the equation $g_{m,n}$ can only be closed if the $g$'s on the RHS are in the set in the state space. The problematic ones are the $g_{m+2,n}$ when $m+n\ge 48$. We assume that $\dot{g}_{m,n}=0$ if $m+n\ge 48$ to close the system. We also tried different orders and truncation schemes and observed equally poor prediction [data not shown], as seen in Fig.~\ref{fig:vanDerPol-Carleman}. 

\noindent{\bf{Data-driven projection.}} To showcase that there exists an approximate Koopman operator acting on a functional subspace in $\mathcal{L}^2\l(\mathbb{R}^n, \rho_0\l(x\r) \dd x \r)$, which reasonably approximates the dynamics, we consider the above polynomial functions up to $m+n\le 4$. This leads to a 10-dimensional functional space. To learn the approximate Koopman operator, we use the numerical solution on a uniform temporal grid with $\delta=0.002$ from $t=0$ to $t=20$ (i.e., 10,000 snapshots of the system). We use the polynomials ($m+n\le 4$) as the dictionary to perform EDMD and learn the one-step approximate Koopman operator in the sub-functional space spanned by the dictionary. The learned $\mathbf{K}_{\delta = 0.002}$ is then used to propagate the initial condition recursively. Both state variables $x$ and $y$, already in this dictionary as $g_{1,0}$ and $g_{0,1}$, are reported in Fig.~\ref{fig:vanDerPol-Carleman}. 

\begin{figure*}[!t]
\centering
    \includegraphics[width=1.0\textwidth]{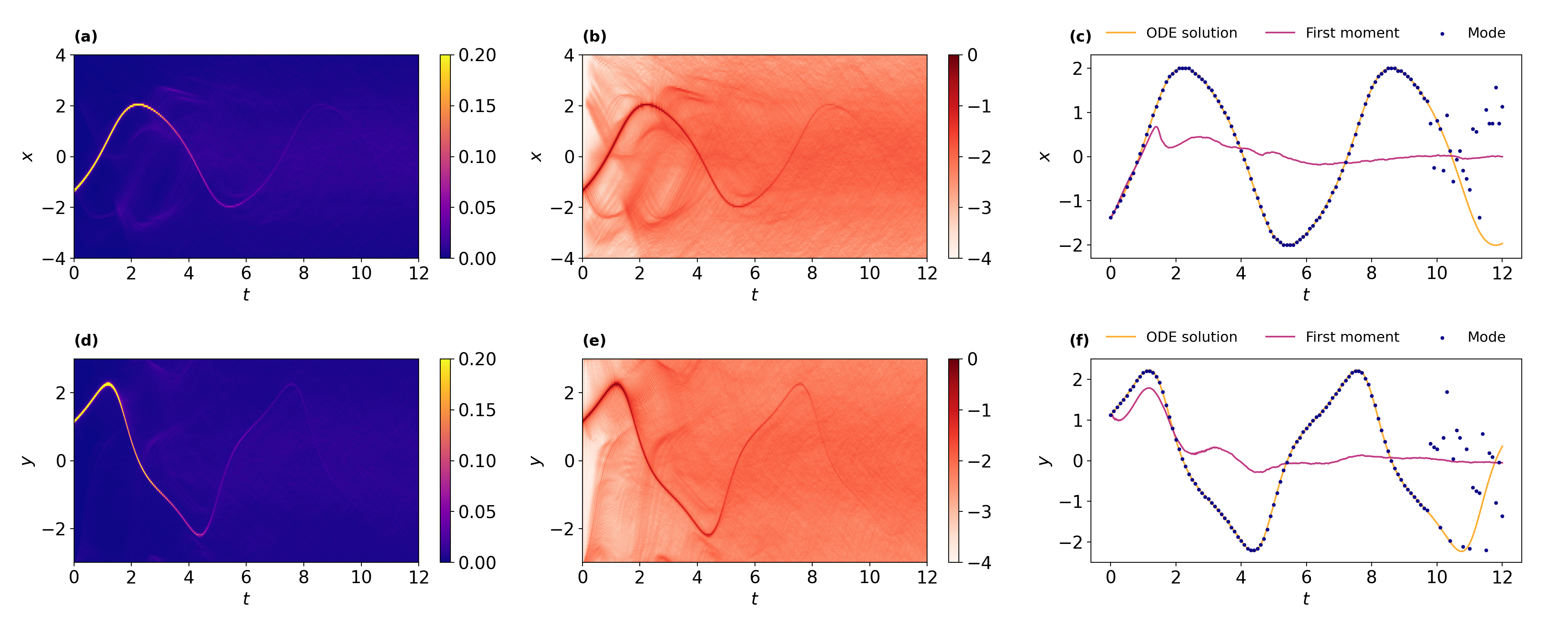}
    \caption{KvN mechanics of the Van der Pol oscillator. }
    \label{fig:vanDerPol-KvN}
\end{figure*}

\noindent{\bf{KvN mechanics.}} We partition the domain $\l(x,y\r) \in \l[-4, 4\r]\times \l[-3,3\r]$ to $128\times 128$ discretized points. The operators $-i\partial_x$ and $-i\partial_y$ are constructed by FFT-based (Fast Fourier Transform) differentiation. We used \texttt{scipy.linalg.expm} to exponentiate the constructed KvN Hamiltonain and obtain finite-time propagator, $\exp\l(-i\mathcal{H}_\text{KvN} \delta\r)$. We chose $\delta=0.01$. The system is evolved forward in time until $t=12$ (total 1,200 time steps). The results are presented in Fig.~\ref{fig:vanDerPol-KvN}. 

\noindent{\bf{Directly solving Liouville equations.}} We partition the domain $\l(x,y\r) \in \l[-4, 4\r]\times \l[-3,3\r]$ to $128\times 128$ discretized points. Denote the coordinates of these discretized points by $(x_j,y_k)$, $j,k \in \l\{1, \ldots, 128\r\}$, we consider a random walk following the below rules on the lattice grid: 
\begin{subequations} \label{eq:RW-VanDerPol}
\begin{align} 
    \l(x_j,y_k\r) \xrightarrow{\frac{\l\vert F_x\l(x_j, y_k\r) \r\vert }{\Delta_x}} \l(x_{j+\text{sign}\l(F_x\l(x_j, y_k\r)\r)}, y_k\r), \\
    \l(x_j,y_k\r) \xrightarrow{\frac{\l\vert F_y\l(x_j, y_k\r) \r\vert }{\Delta_y}}\l(x_j, y_{k+\text{sign}\l(F_y\l(x_j, y_k\r)\r)}\r).
\end{align}
\end{subequations}
Here, $F_x$ and $F_y$ are the flow defined in \eqref{eq:ODEvanDerPol}. The transition rates are annotated in above the arrows. We also truncate the reactions so that the random walker is always confined on the $128\times 128$ lattice.
The joint probability distribution, $p_{i,j}\l(t\r)=\mathbb{P}\l\{\text{The random walker is at }x_j, y_k \text{ at time } t\r\}$, $i,j\in \l\{1, \ldots 128\r\}$ follows a CME, which is our object of interest. By stacking up the probabilities into a $16384\times 1$ column vector $\mathbf{p}(t)$, the CME is $\dot{\mathbf{p}}\l(t\r) = \mathcal{L}_{\Delta }^\dagger \cdot \mathbf{p}\l(t\r)$. We then solve the operator $\exp \l(\delta \mathcal{L}_\Delta^\dagger\r)$ by exponentiation of the matrix, using \texttt{scipy.linalg.expm}, for forward propagating the  initial density. Similar to the setting in the KvN mechanics, we chose $\delta=0.01$ and propagate the initial state 1,200 steps. 

\begin{figure*}[!t]
\centering
    \includegraphics[width=1.0\textwidth]{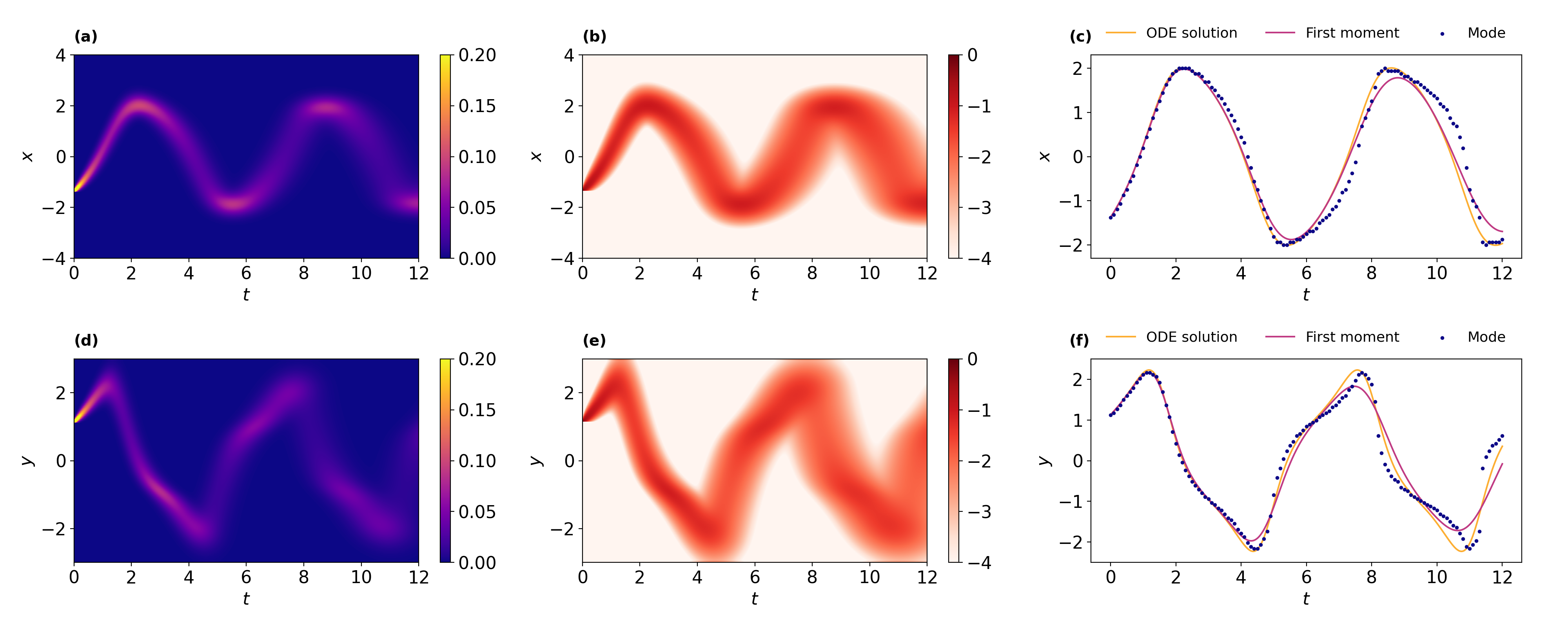}
    \caption{CME approximating the Liouville dynamics of the Van der Pol oscillator. }
    \label{fig:vanDerPol-Liouville}
\end{figure*}

The results, presented in Fig.~\ref{fig:vanDerPol-Liouville}, showed a clear signature of stochastic phase diffusion where the probability density is dispersed near the stable limit cycle. For this system and with this parametrization ($\mu=0.5$), the mode of the KvN mechanics (Figs.~\ref{fig:vanDerPol-KvN}(c) and (f)) is almost identical to the true solution and is more accurate than that of the CME (Figs.~\ref{fig:vanDerPol-Liouville}(c) and (f)), which suffered from phase diffusion. However, we observed that the first moments of the CME are more reliable than that of the KvN mechanics, which suffers from the Gibbs-like phenomena (Figs.~\ref{fig:vanDerPol-KvN})(a-b) and (d-e).

\newcommand{\Fx}{\l(F_x\r)_{i,j}}
\newcommand{\Fy}{\l(F_y\r)_{i,j}}
\newcommand{\rh}{\rho\l(x_i, y_j, t\r)}

Following the approach that led to (\ref{eq:FOUphi2}), a first-order upwind discretization of Liouville dynamics for the Van der Pol oscillator may be written as
\begin{multline}
    \label{eq:FOUvanderpol}
    \rho_{j,k}^{n+1} = \rho_{j,k}^n -
            \frac{\delta}{\Delta_x} \left[\left(F_x \rho\right)^n_{j+1/2,k} - \left(F_x \rho\right)^n_{j-1/2,k}\right]
            \\-
            \frac{\delta}{\Delta_y} \left[\left(F_y \rho\right)^n_{j,k+1/2} - \left(F_y \rho\right)^n_{j,k-1/2}\right],
\end{multline}
where the interface fluxes are defined by
\begin{subequations}
\begin{align}
    \left(F_x \rho\right)^n_{j+1/2,k} ={}& \l(\phi_R\r)^n_{j+1/2,k} + \l(\phi_L\r)^n_{j+1/2,k},  \label{eq:2DFOU-head}\\
    \l(\phi_R\r)^n_{j+1/2,k} :={}& \l\{\begin{array}{cl}
         \frac{\Fx}{\Delta_x} \rh & \text{ if } \Fx>0  \\
         0 & \text{ else,} 
    \end{array}\r.\\
    \l(\phi_L\r)^n_{j+1/2,k} :={}& \l\{\begin{array}{cl}
         \frac{\l(F_x\r)_{i+1,j}}{\Delta_x} \rho\l(x_{i+1}, y_j, t\r) & \text{ if } \l(F_x\r)_{i+1,j} < 0   \\
         0 & \text{ else,} 
    \end{array}\r.
\end{align}
where $\phi_R$ and $\phi_L$ are right- and left-passing fluxes crossing the same interface $\l(j+1/2, k\r)$, $\Fx:=F_x\l(x_i, y_j\r)$ and
\begin{align}
    \left(F_y \rho\right)^n_{j,k+1/2} ={}& \l(\phi_U\r)^n_{j,k+1/2} + \l(\phi_D\r)^n_{j,k+1/2},  \\
    \l(\phi_U\r)^n_{j,k+1/2} :={}& \l\{\begin{array}{cl}
         \frac{\Fy}{\Delta_y} \rh & \text{ if } \Fy>0  \\
         0 & \text{ else,} 
    \end{array}\r.\\
    \l(\phi_D\r)^n_{j,k+1/2} :={}& \l\{\begin{array}{cl}
         \frac{\l(F_y\r)_{i,j+1}}{\Delta_y} \rho\l(x_{i}, y_{j+1}, t\r) & \text{ if } \l(F_y\r)_{i,j+1} < 0   \\
         0 & \text{ else,} 
    \end{array}\r. \label{eq:2DFOU-tail}
\end{align}
\end{subequations}
where $\phi_U$ and $\phi_D$ are upward- and downward-passing fluxes crossing the same interface $\l(j, k+1/2\r)$, and $\Fy:=F_y\l(x_i, y_j\r)$. Our choices here for the finite-volume discretization, Eqs.~\eqref{eq:2DFOU-head} to \eqref{eq:2DFOU-tail}, led to an evolutionary equation \eqref{eq:FOUvanderpol} that matches the CME describing the probability distribution of the random walk exactly  \eqref{eq:RW-VanDerPol}.

\section{Finite-time Blowup of Carleman Linearization} \label{app:carleman}

Carleman linearization has been used in quantum computing
in order to solve nonlinear systems. The error bounds derived by Forets
and Pouly \cite{forets_explicit_2017} may justify
the use of Carleman, but one must keep in mind the time interval over
which the error bounds are valid, specifically \cite[Eq.~(25)]{forets_explicit_2017}.
Of course, one possibility is that the time interval derived by Forets
and Pouly is an estimate and too pessimistic and solutions may
be accurate for longer times. However, in this appendix,
we show that for a simple ODE, the time interval derived by Forets
and Pouly is exact.  After the maximum time is exceeded, we show that solutions become unbounded in the limit of infinite truncation order.

Consider the scalar, real-valued ODE
\begin{equation}
\frac{dx}{dt}=-rx-\varepsilon x^{2}\,,\label{eq:scalar}
\end{equation}
with the initial condition $x(0)=x_{0}$. This
ODE may be viewed as a nonlinear $\varepsilon$-perturbation of a
linear ODE. Below, we show that the solution from a truncated Carleman
linearization, with exact time integration, is in the form of a power
series in $\varepsilon$. As such, Carleman solutions are bounded
only when the solution is within the radius of convergence of the
power series, which is more restrictive than conditions for bounded
exact solutions. When outside the convergence radius, Carleman solutions
blow up in finite time, even with
exact time integration. This result indicates that changing the time-integration
method will have a modest effect on the restrictive stability conditions
for Carleman applied to more complicated problems, such as the application to Burgers'
equation in \cite{liu2021efficient}. Another viewpoint is that
in order for Carleman to be accurate for long-time simulations of \eqref{eq:scalar} requires small $|\varepsilon|$. The conditions on $\varepsilon$ will be made precise below.

\noindent{\bf{Properties of the Exact Solution.}}
For $|r|>0$, Eq.~\eqref{eq:scalar} has 1 stable and 1 unstable fixed
point, which we denote as $x_{s}$ and $x_{u}$, respectively. Their
values are given by
\begin{equation}
x_{s}=\frac{1}{\varepsilon}\max\left(0,-r\right)\qquad x_{u}=\frac{1}{\varepsilon}\min\left(0,-r\right)\label{eq:fixed_points}
\end{equation}
Whenever $r\equiv0$, the point $x_{s}=x_{u}=0$ is a stable point for $\varepsilon>0$ and unstable for $\varepsilon<0$.

The exact solution of \eqref{eq:scalar} is given by
\begin{equation}
x(t)=\frac{x_{0}e^{-rt}}{1+R\left(1-e^{-rt}\right)}\,.\label{eq:exact_scalar}
\end{equation}
where $R=x_{0}\varepsilon/r$. Note that the numerator $x_{0}e^{-rt}$
is the solution for $\varepsilon=0$, while the $R$-term represents
the contribution from nonlinearity. For $r\equiv0$, this solution
reduces to
\begin{equation}
x(t)=\frac{x_{0}}{1+\varepsilon x_{0}t}\,.\label{eq:exact_scalar_r0}
\end{equation}

In order for the exact solution
to remain bounded for all time, we require
\begin{itemize}
    \item $R>1$ for $r>0$,
    \item $R\le0$ for $r < 0$, and
    \item $\varepsilon x_0 > 0$ for $r=0$.
\end{itemize}
Under these conditions, the exact solution remains finite and monotonically approaches $x=x_{s}$ as $t\rightarrow\infty$.

\noindent{\bf{Carleman Linearization.}}
The Carleman linearization of \eqref{eq:scalar} solves 
\begin{equation}
\frac{\dd x^{m}}{\dd t}=-mrx^{m}-m\varepsilon x^{m+1}\,,\label{eq:carl_scalar_ODE}
\end{equation}
for $1\le m\le M$. This system may be written as the linear system
\begin{equation}
\frac{\dd \hat{y}}{\dd t}=\hat{H}\hat{y}\,,\label{eq:carl_scalar_vectorform}
\end{equation}
where each component of $\hat{y}$ is given by $\hat{y}_{m}=x^{m}$
and $\hat{H}\in\mathbb{R}^{M\times M}$.
We assume the closure
\begin{equation}
\frac{\dd x^{M}}{\dd t}=0\,,\label{eq:closure}
\end{equation}
which is satisfied if $\hat{y}$ is at a fixed point. It's
tempting then to set the initial condition of the component $\hat{y}_M$ as
\begin{equation}
\hat{y}_{M}(0)=x_{s}^{M}\,.\label{eq:closure_fp}
\end{equation}
However, for large $M$, our numerical experiments found no improvement when using \eqref{eq:closure_fp}, compared to setting the initial condition for all components as simply
\begin{equation}
\hat{y}_{m}(0)=x_{0}^{m}\,,
\end{equation}
for $1\le m\le M$. This is the initial condition we used for the remainder of this study.

Other closures than \eqref{eq:closure} should
be considered for future work, but we were unable to find a closure
that qualitatively changes the conclusions of this study. We suspect
that the reason the closure has little impact is that if $x^{M}$
has a large impact on the Carleman solution, then either $M$ is too small or
the $M\rightarrow\infty$ solution is unbounded.

We now derive the exact solution of the Carleman linearization Eq.~\eqref{eq:carl_scalar_vectorform}. First, we express the matrix $\hat{H}= -r \hat{D} - \varepsilon \hat{A}$, with $\hat{D}:=\text{diag}\l(1, 2, \ldots M\r)$ and $\l[\hat{A}\r]_{i,j}:=i \delta_{j, i+1}$. Then, we perform a change of variable 
\begin{equation}
  \hat{z}\l(t\r):= \hat{\Theta} \cdot \hat{y}\l(t\r),
\end{equation}
with a defined matrix $\hat{\Theta}:=\text{diag}\l(e^{rt}, e^{2rt}, \ldots e^{Mrt}\r)$. Apply $\dd/\dd t$ and chain rule to $\hat{z}(t)$, we obtained
\begin{equation}
    \frac{\dd}{\dd t} \hat{z}(t) = -\varepsilon \hat{\Theta} \cdot \hat{A}\cdot  \hat{\Theta}^{-1} \cdot \hat{z}\l(t\r) = - \varepsilon e^{-rt} \hat{A} \cdot \hat{z}\l(t\r). 
\end{equation}
To solve the above equation, it is useful to perform a non-homogeneous scaling of the time $t$ by defining $\tau := \varepsilon \l(1-e^{-rt }\r)/r$. As $\dd \tau / \dd t = \varepsilon e^{-rt}$, we can express the above equation into the canonical form
\begin{equation}
    \frac{\dd}{\dd \tau} \hat{z}(\tau ) = - \hat{A} \cdot \hat{z}\l(\tau\r),
\end{equation}
whose solution is simply
\begin{equation}
    \hat{z}\l(\tau\r) = e^{-\tau \hat{A}}\cdot \hat{z}\l(0\r) = \sum_{m=0}^{\infty} \frac{\l(-\tau A\r)^{m}\cdot \hat{z}\l(0\r)}{m!}.
\end{equation}
It is elementary to show the $(i,j)$-entries of $\hat{A}^m$ is $\delta_{j, i+m} \l(i+m-1\r)!/\l(i-1\r)! $ for $m < M$. Also, $\hat{A}^m=0$ for $m\ge M$. Thus, the above solution can be further simplified
\begin{equation}
    \hat{z}\l(\tau\r) =\sum_{m=0}^{M-1} \frac{\l(-\tau \hat{A}\r)^{m}\cdot \hat{z}\l(0\r)}{n!}. 
\end{equation}
As we only care about the $x(t)$ which can be uniquely determined by the first component of $\hat{z}$,
\begin{align}
    \hat{z}_1\l(\tau\r) ={}&\sum_{m=0}^{M-1} \sum_{k=1}^M \frac{\l(-\tau\r)^m}{m!}  \l[\hat{A}^{m}\r]_{1,k} \hat{z}_k\l(0\r) \nonumber\\
    ={}& \sum_{m=0}^{M-1} \l(-\tau\r)^m \hat{z}_{1+m}\l(0\r). 
\end{align}
In the last equality, we used the identity that the $(1,k)$-entry of $\hat{A}^{m}$ is $\l(m!\r) \delta_{k, 1+m}$. 
Finally, plugging back $-\tau = r\l(e^{-rt}-1\r)/\varepsilon$, $x(t)=\hat{y}_1\l(t\r) = e^{-rt} \hat{z}_1\l(t\r)$, and $\hat{z}_{m+1}\l(0\r) = \hat{y}_{m+1}\l(0\r)= x_0^{m+1}$, we obtained the analytic solution to Eq.~\eqref{eq:carl_scalar_vectorform}:
\begin{equation}
x(t)=x_{0}e^{-rt}\sum_{m=0}^{M-1}\left[\left(e^{-rt}-1\right)R\right]^{m}\,.\label{eq:carleman_solution}
\end{equation}
Recognize that for $\vert R \vert$ small (i.e., either $\vert \varepsilon \vert$ or $\vert x_0\vert$ is small, or $\vert r \vert $ is large),
the following term in the exact solution \eqref{eq:exact_scalar} may be evaluated
as a power series:
\begin{equation}
\frac{1}{1+R\left(1-e^{-rt}\right)}=\sum_{m=0}^{\infty}\left[\left(e^{-rt}-1\right)R\right]^{m}\,.\label{eq:eps_series}
\end{equation}
We remark that a similar analysis can be performed to the special case $r=0$; in this case, $\hat{z}=\hat{y}$ and $\tau=\varepsilon t$ which leads to 
\begin{equation}
x(t)=x_{0}e^{-rt}\sum_{m=0}^{M-1}\left(\varepsilon x_0 t \right)^{m}\,,\label{eq:carleman_solution_r0}
\end{equation}
which can also be considered as an expansion of Eq.~\eqref{eq:exact_scalar_r0} with respect to $\varepsilon x_0 t $. Therefore, the truncated Carleman linearization of \eqref{eq:scalar}
may be viewed as a truncation of the $\varepsilon$-series expansion
of the exact solution. Truncation of Carleman introduces error, but
more importantly, the series \eqref{eq:eps_series} may not converge
and result in blowup as $M\rightarrow\infty$. The convergence behavior
of the series depends primarily on the sign of $r$ and each case
is discussed below.

\noindent{\bf{Conditions for Bounded Solutions, $\boldsymbol{r>0}$.}}
In this case, we find that in order
to obtain bounded solutions for all $t>0$, the Carleman linearization
is more restrictive than the exact solution. To show this, recall that
a power series such as \eqref{eq:eps_series} has a convergence radius
of 1, so convergence requires that
\begin{equation}
\left|\left(e^{-rt}-1\right)R\right|<1\,,\label{eq:conv_crit}
\end{equation}
for all $t>0$. The maximum value of the left-hand side is attained
as $t\rightarrow\infty$ and results in the condition
\begin{equation}
-1<R<1\,.\label{eq:r_cond}
\end{equation}
This condition is analogous to that found by \cite{liu2021efficient}
for Carleman applied to Burgers' equation.
The first inequality is also required by the exact solution, while
$R<1$ is an additional restriction on the Carleman linearization.
When \eqref{eq:r_cond} is not satisfied, the Carleman
solution will blow up in finite time as $M\rightarrow\infty$. The time past which blowup occurs
may be found from \eqref{eq:conv_crit}, which gives
\begin{equation}
t_{\text{}}>-\frac{1}{r}\log\left(1-|R|^{-1}\right)\,.\label{eq:tblowup_rpos}
\end{equation}
This time is identical to that found for the validity of the error
bound derived by \cite[eq.  (25)]{forets_explicit_2017}.

\noindent{\bf{Conditions for Bounded Solutions, $\boldsymbol{r<0}$}.}
In this case, we find
that except for the initial condition $x_{0}=0$, all Carleman solutions
blow up in finite time. Because $e^{-rt}\ge1$,
the series convergence condition \eqref{eq:conv_crit} reduces to
\[
e^{-rt}-1<|R|^{-1}
\]
so that the series blows up for
\begin{equation}
t>-\frac{1}{r}\log\left(1+|R|^{-1}\right)\,,\label{eq:tblowup_rneg}
\end{equation}
for all $|R|>0$. Note that \cite{forets_explicit_2017} did not consider
the case equivalent to $r<0$; in their notation, they assumed $\mu(F_{1})\le0$.

\noindent{\bf{Conditions for Bounded Solutions, $\boldsymbol{r=0}$.}}
In this case, one can show that eq. \eqref{eq:carleman_solution} reduces
to 
\begin{equation}
x(t)=x_{0}\sum_{m=0}^{M-1}\left(-\varepsilon x_{0}t\right)^{m}\,.\label{eq:phi2_carleman_exact-1}
\end{equation}
The series blows up as $M\rightarrow\infty$ for
\begin{equation}
t>\frac{1}{\left|x_{0} \varepsilon\right|}\,.\label{eq:tr0}
\end{equation}
This result is consistent with the numerical results presented in Fig. \ref{fig:carleman}
 (where $x_{0}=\varepsilon=1)$ and the error
bound derived in \cite[eq.  (25) with $\mu(F_1)=0$.]{forets_explicit_2017}.

\begin{figure}[!t]
\begin{centering}
\includegraphics[width=1.0\columnwidth]{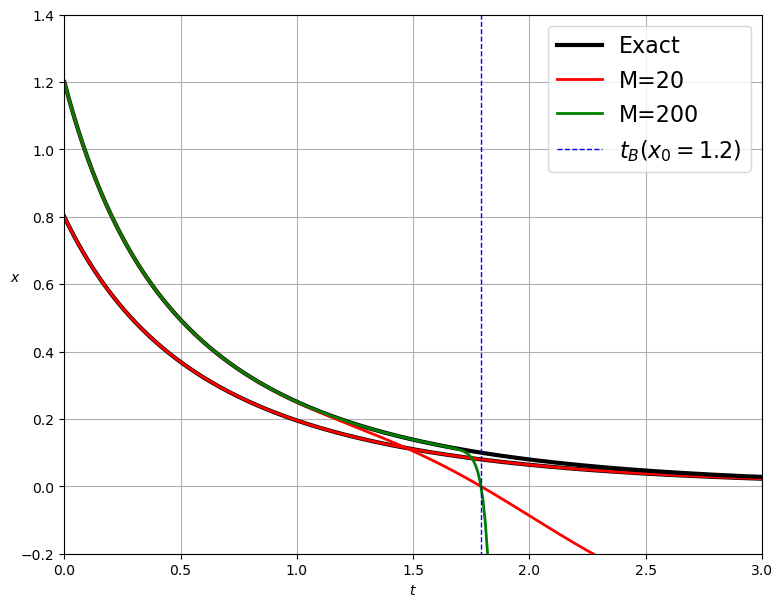}
\par\end{centering}
\caption{Solutions for $r=1$, $\varepsilon=1$, and two initial conditions: $x_{0}=R=0.8$ ($|R|<1$; Carleman bounded) and $x_{0}=R=1.2$ ($R>1$;
Carleman blows up for $t>t_{B}$ as $M\rightarrow\infty$). For $x_{0}=0.8$,
only Carleman($M=20$) is shown and is coincident with the exact solution.
The stable fixed point is $x_{s}=0$, and the value of $t_{B}$ was
computed from the right-hand side of \eqref{eq:tblowup_rpos}.\label{fig:r1}}
\end{figure}

\begin{figure}[!t]
\begin{centering}
\includegraphics[width=1.0\columnwidth]{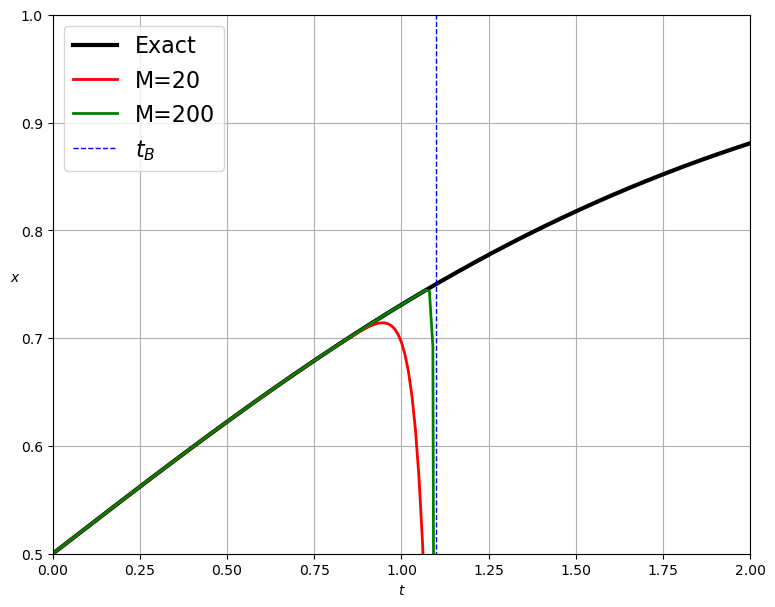}
\par\end{centering}
\caption{Solutions for $r=-1$, $\varepsilon=1$, and $x_{0}=0.5$. Note that
for $r<0$ and any initial condition, Carleman blows up for $t>t_{B}$
as $M\rightarrow\infty$. The stable fixed point is $x_{s}=1$, and
the value of $t_{B}$ was computed from the right-hand side of \eqref{eq:tblowup_rneg}.\label{fig:rm1}}
\end{figure}

\noindent{\bf{Summary of Carleman Blowup}.}
The results of this appendix are summarized in Table \ref{tab:Behavior-of-Carleman}.
The terminology here is as follows:
\begin{itemize}
\item \emph{Bounded}: The solution is bounded for all $t\ge0$.
\item \emph{Unbounded}: The solution is unbounded at one or more times.
In these cases, both the exact and Carleman solution are unbounded,
so the differences between these solutions were not investigated.
\item \emph{Blowup}: The Carleman solution is unbounded for all times greater than
that specified, as the truncation order $M\rightarrow\infty$.
\end{itemize}
\begin{table}[H]
\begin{centering}
{\renewcommand\arraystretch{1.75}%
\begin{tabular}{|l|l|l|l|}
\hline 
\multicolumn{2}{|l|}{\textbf{Case}} & \textbf{Exact} & \textbf{Carleman}\tabularnewline
\hline 
\multirow{3}{*}{$r>0$} & $|R|<1$ & Bounded & Bounded\tabularnewline
\cline{2-4} \cline{3-4} \cline{4-4} 
 & $R\le-1$ & Unbounded & Unbounded\tabularnewline
\cline{2-4} \cline{3-4} \cline{4-4} 
 & $R\ge1$ & Bounded & \textcolor{red}{Blowup for $t_{\text{}}>-\frac{1}{r}\log\left(1-|R|^{-1}\right)$}\tabularnewline
\hline 
\multirow{2}{*}{$r=0$} & $\varepsilon x_0 <0$ & Unbounded & Unbounded\tabularnewline
\cline{2-4} \cline{3-4} \cline{4-4} 
 & $\varepsilon x_0> 0$ & Bounded & \textcolor{red}{Blowup for $t>\frac{1}{\left|x_{0} \varepsilon\right|}$}\tabularnewline
\hline 
\multirow{2}{*}{$r<0$} & $R>0$ & Unbounded & Unbounded\tabularnewline
\cline{2-4} \cline{3-4} \cline{4-4} 
 & $R\le0$ & Bounded & \textcolor{red}{Blowup for $t>-\frac{1}{r}\log\left(1+|R|^{-1}\right)$}\tabularnewline
\hline 
\end{tabular}}
\par\end{centering}
\caption{Behavior of Exact and Carleman solutions of Eq. \eqref{eq:scalar}.
The deviation of Carleman from the Exact behavior is highlighted in
red. \label{tab:Behavior-of-Carleman}}
\end{table}
Sample exact solutions of \eqref{eq:scalar} and \eqref{eq:carl_scalar_ODE} are shown
in Figs. \ref{fig:r1} and \ref{fig:rm1}.

Finally, note that \cite[supplimentary material]{liu2021efficient} showed that when solving a system of two copies of \eqref{eq:scalar}, for $r=1$ and $\varepsilon < 0$, quantum exponential speedup is not possible for $R < -1$ (corresponding to $R>1$ for their definition of $R$).  Their result is general and does not require using Carleman linearization. But exact solutions for this case are unbounded for large time. Future work should investigate whether their result extends to cases that have bounded exact solutions. 

\end{document}